\documentclass[pra,reprint]{revtex4-1}
\usepackage{graphicx}
\usepackage{enumitem}
\usepackage{amsmath}
\newcommand{\ket}[1]{|{#1}\rangle}

\usepackage{amsfonts}

\usepackage[usenames,dvipsnames]{xcolor}
\usepackage[colorlinks=true,citecolor=Cerulean,linkcolor=RubineRed,urlcolor=Cerulean]{hyperref}

\begin{document}

\title{Mean-field phases of an ultracold gas in a quasicrystalline potential}

\author{Dean Johnstone}
\email{dj79@hw.ac.uk}
\author{Patrik \"Ohberg}
\author{Callum W. Duncan}
\email{cd130@hw.ac.uk}
\affiliation{SUPA, Institute of Photonics and Quantum Sciences,
	Heriot-Watt University, Edinburgh, EH14 4AS, UK}

\begin{abstract}
The recent experimental advancement to realise ultracold gases scattering off an eight-fold optical potential [Phys. Rev. Lett. \textbf{122}, 110404 (2019)] heralds the beginning of a new technique to study the properties of quasicrystalline structures. Quasicrystals possess long-range order but are not periodic, and are still little studied in comparison to their periodic counterparts. Here, we consider an ultracold bosonic gas in an eight-fold symmetric lattice and assume a toy model where the atoms occupy the ground states of the local minima of the potential. The ground state phases of the system are studied, with particular interest in the local nature of the phases. The usual Mott-insulator, density wave, and supersolid phases of the standard and extended Bose-Hubbard model are observed. For non-zero long-range interactions, we find that density wave states can spontaneously break the eight-fold symmetry, and may even possess no rotational symmetry. We find the local variation in the number of nearest neighbours to play a vital role in the phase transitions, local structure, and global symmetries of the ground states. This variation in the number of nearest neighbours is not a unique property of the considered eight-fold lattice, and we expect our results to be generalisable to any quasicrystalline potential where there are only small position dependent variations in the site energy, tunnelling and interactions.
\end{abstract}

\maketitle

\section{Introduction}

Quasicrystals are a state of matter that have long-range order but are not periodic \cite{Shechtman1984,Levine1984,steurer2018}. Their order can be considered to arise from incommensurate projections of higher-dimensional periodic crystals \cite{Lang2012,Kraus2013}, or due to a continous tiling of space with discrete unit cells  \cite{penrose1974}. Over the last few decades, quasicrystals have been studied in condensed matter systems \cite{jenks1998,steurer2018,Louzguine2008}, photonics \cite{Jin2012,vardeny2013}, twisted bilayer graphene \cite{Yao2018}, the Gross-Pitaevskii equation with solitons \cite{Sakaguchi2006}, and as an emergent phase in ultracold dipolar gases \cite{Gopalakrishnan2013}. There has also been a significant amount of research into the case of one-dimensional disordered, quasirandom and/or incommensurate systems that are related to quasicrystals \cite{Lu1986,HIRAMOTO1992,Fallani2007,roati2008,Edwards2008,Gadway2011,Errico2014,Singh2015,schreiber2015,Hu2016,Duncan2017}. It was also shown recently that general quantum Hamiltonians, including those of quasicrystals, can be solved exactly by extending the problem to a higher-dimensional space, which is called a superspace \cite{valiente2019}. 

Lately the consideration of two-dimensional quasicrystalline optical lattices for atomic gases has been gaining traction \cite{Guidoni1997,Sanchez2005,Jagannathan2013,Hou2018,Corcovilos2019}, resulting in the first experimental work considering the scattering of an atomic gas from a quasicrystalline optical potential \cite{Viebahn2019}. An interesting extension to consider is the adiabatic loading of an atomic gas into a quasicrystalline optical lattice, which is currently possible for relatively weak lattice depths \cite{viebahn2018}. However, with current experimental technologies, potentially combined with enforced adiabatic loading schemes \cite{Zhou2018}, we expect the adiabatic loading of an ultracold gas into a quasicrystalline optical lattice to be experimentally achievable in the near future. With the realisation of Bose-Hubbard models in quasicrystal lattices, potentially important questions on the relation between quasiperiodic order and randomness could be probed \cite{Viebahn2019,Khemani2017}.

Ground state phases for ultracold atoms in an optical lattice were first studied in the standard Bose-Hubbard model \cite{Fisher1989,Freericks1994,greiner2002,Zwerger2003,bakr2010}, in which the well-known Mott-insulator to superfluid transition arises. However, by introducing long-range interactions, density wave and supersolid phases can also be observed \cite{Tomasz2012,Rossini2012,Ohgoe2012,lewenstein2012}. These phases spontaneously break the translational symmetry of a lattice and can destroy the previously present superfluid and Mott-insulating domains. Long-range interactions can be introduced by the use of dipolar atomic species \cite{Trefzger2011,Aikawa2012,Lu2012,baier2016}, or can be induced by light-matter interactions \cite{Caballero2016,Dogra2016,Niederle2016,landig2016,Flottat2017}.

In this work, we will consider the case of bosons in a lattice geometry based on an eight-fold symmetric optical lattice used in a recent ultracold gas experiment \cite{Viebahn2019}. We will assume a toy model for this system, with the atoms occupying the states associated to the local minima. The ground state phases of both standard and extended Bose-Hubbard models for the quasicrystalline lattice are studied by a static Gutzwiller mean-field approach using a self-consistent loop. First, we will investigate the case of a homogeneous structure on the quasicrystal, with constant tunnelling and interaction strengths across the lattice. This allows us to study how variations in the local number of nearest neighbours throughout the lattice will impact the ground state phases. We will then finish this work by considering the position-dependent tunnelling and interactions within the quasicrystalline lattice and confirm that the phases are present in this scenario.

\section{The quasicrystalline lattice}

\begin{figure}[t]
	\centering
	\includegraphics[width=0.98\linewidth]{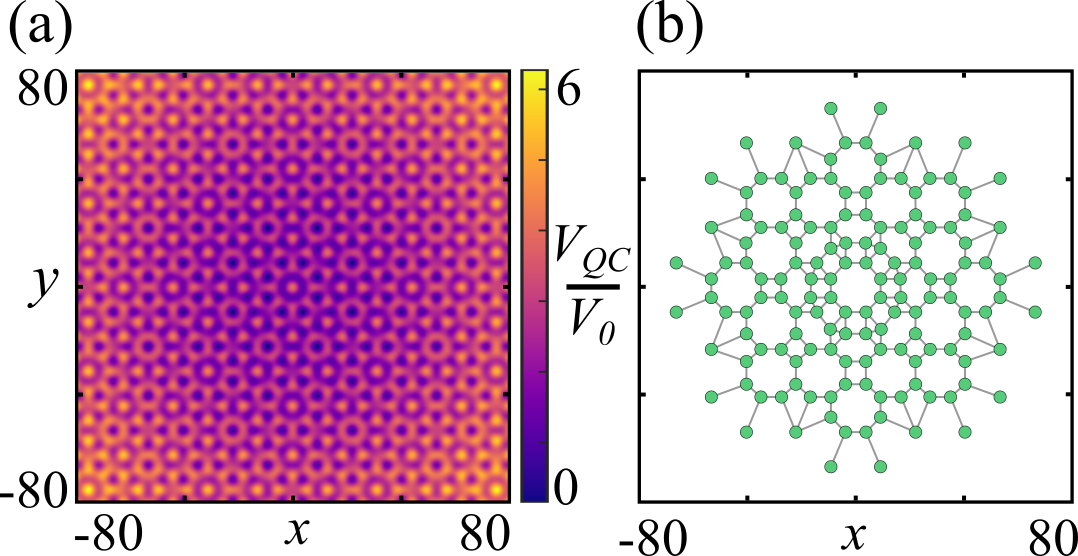}
	\caption{The eight-fold symmetric lattice. (a) The optical potential of Eq.~\eqref{eq:Lattice} with a weak harmonic trap $V_H =  0.0004V_0 \lambda_{\mathrm{latt}}^{-2}$. (b) Positions of lattice sites when taking cut-off value of $V_{QC}/V_0=1$ in (a), bonds are shown for nearest neighbour couplings. Positions are in units of $\lambda_{\mathrm{latt}}$ and $V_0 = 0.8E_R$.}
	\label{fig:lattice}
\end{figure}

For two-dimensional systems, a crystal can have only two-, three-, four- and/or six-fold rotational symmetry \cite{Landau1958}. As already stated, we will consider a lattice based on an optical potential which is eight-fold symmetric, which is defined by
\begin{equation}
V_{QC} (\mathbf{r}) = V_0 \sum_{i=1}^{4} \cos^2 \left(\frac{\mathbf{\hat{G}_i}}{2} \cdot \mathbf{r} \right),
\end{equation}
where $V_0$ defines the lattice strength, $\mathbf{r}$ the two-dimensional position (in units of $\lambda_{\mathrm{latt}}$), and the reciprocal lattice vectors are
\begin{equation}
\begin{aligned}
\mathbf{\hat{G}_1} = \begin{pmatrix}
1 \\ 0
\end{pmatrix},
\mathbf{\hat{G}_2} = \begin{pmatrix}
0 \\ 1
\end{pmatrix}, & \\
\mathbf{\hat{G}_3} = \frac{1}{\sqrt{2}}\begin{pmatrix}
1 \\ 1
\end{pmatrix},
\mathbf{\hat{G}_4} = \frac{1}{\sqrt{2}}\begin{pmatrix}
-1 \\ 1
\end{pmatrix}. &
\end{aligned}
\end{equation}
This lattice potential can be realised by interfering four mutually non-interfering one-dimensional optical lattices and was recently realised experimentally \cite{Viebahn2019,viebahn2018}. We will also consider, in part, the influence of a harmonic trap required to trap the atomic cloud. We will account for the harmonic term in the full lattice potential, which we define as
\begin{align}
V_{\mathrm{latt}}(\mathbf{r}) = V_{QC} (\mathbf{r}) + \frac{1}{2} V_{H} \mathbf{r}^2,
\label{eq:Lattice}
\end{align}
where $V_H$ is a tunable harmonic strength. The form of the lattice potential is shown in units of $V_0$ in Fig.~\ref{fig:lattice}a. We consider a cut-off of $V_{QC}/V_0 = 1$ to obtain the toy model of a finite lattice of the quasicrystal. This cut-off is equivalent to the setting of an energy (temperature) of the ultracold gas and/or the depth of the lattice. In this case, the gas occupies minima in the lattice potential where $V_{QC}/V_0 < 1$, giving the finite lattice shown in Fig.~\ref{fig:lattice}b. 

In the true physical system the atoms would usually be loaded into a single band (or multiple bands) of the quasicrystalline potential which would require knowledge of the band structure and the positions of energy gaps. With the cut-off approach used in this work, we will consider the case that the ultracold gas has been cooled sufficiently to sit in the local ground states of the potential minima for the quasicrystalline optical lattice. This is a toy model which we will use to give a qualitative picture of the ground state phases.
 
\section{Model}
\label{sec:Model}

An interacting ultracold gas trapped in an optical lattice has the Hamiltonian 
\begin{equation} \label{eq_fieldH}
\begin{aligned}
\mathcal{\hat{H}} = \int d\mathbf{\mathbf{r}} \hat{\Psi}^\dagger (\mathbf{r}) \bigg( -\dfrac{\hbar^2}{2m}\nabla^2 + V_{\mathrm{latt}}(\mathbf{r}) - \mu \bigg) \hat{\Psi} (\mathbf{r}) \\ + \, \dfrac{1}{2} \int d\mathbf{r} d \mathbf{r^\prime} \hat{\Psi}^\dagger (\mathbf{r}) \hat{\Psi}^\dagger (\mathbf{r^\prime}) V_{\mathrm{int}}(\mathbf{r},\mathbf{r^\prime}) \hat{\Psi} (\mathbf{r}) \hat{\Psi} (\mathbf{r^\prime}),
\end{aligned}
\end{equation}
where $V_{\mathrm{int}}(\mathbf{r},\mathbf{r^\prime})$ is the two-atom interaction potential, $\mu$ the chemical potential, $m$ the mass,  and $ \hat{\Psi} (\mathbf{r}) \, (\hat{\Psi}^\dagger (\mathbf{r})) $ are the bosonic annihilation (creation) field operators. We will define the interaction potential to be governed by a short- and long-range term, i.e.
\begin{equation}
V_{\mathrm{int}}(\mathbf{r},\mathbf{r^\prime}) = g\delta(\mathbf{r}-\mathbf{r^\prime}) + \gamma h(\mathbf{r},\mathbf{r^\prime}),
\end{equation}
where $g$ and $\gamma$ are the short- and long-range strengths respectively. The short-range interaction is mediated by a contact interaction, given by a delta function. The long-range interaction is mediated by a general term $h(\mathbf{r},\mathbf{r^\prime})$, which we will consider to have a position dependence of $1/|\mathbf{r}-\mathbf{r^\prime}|^3$, which is consistent with that of dipolar atomic gases \cite{Trefzger2011}.

It is well-known that the tight-binding limit of Hamiltonian~\eqref{eq_fieldH} results in a Bose-Hubbard model \cite{Jaksch1998}. If we retain two-site interactions while taking the tight-binding limit we obtain the extended Bose-Hubbard model which has a Hamiltonian of \cite{Dutta2015}
\begin{equation} \label{eq_ebh}
\begin{aligned}
\hat{H} = & - \sum\limits_{\langle i,j \rangle} J_{ij} \hat{b}_{i}^\dagger \hat{b}_{j} + \sum\limits_{i} \dfrac{U_i}{2} \hat{n}_i(\hat{n}_i - 1) \\ & + \sum\limits_{i} \left( \epsilon_i - \mu \right) \hat{n}_i  + \, \sum\limits_{\langle i,j \rangle} V_{ij} \hat{n}_i\hat{n}_j,
\end{aligned}
\end{equation}
where $i(j)$ label the lattice sites, $\langle i,j \rangle$ denotes the sum over all nearest neighbours, $ \hat{b}_i \, (\hat{b}^\dagger_i) $ are the individual bosonic annihilation (creation) operators and $\hat{n}_i = \hat{b}_{i}^\dagger \hat{b}_{i}$ is the number operator. The extended Bose-Hubbard model is the simplest tight-binding Hamiltonian for atoms that have long-range interactions, and it assumes that the additional two-site terms (density-dependent tunnelling and pair tunnelling) are small compared to the single-particle tunnelling. We have limited the summations to nearest neighbours (and the reciprocal term) as the strength for each term falls off quickly with the separation between sites. In Hamiltonian~\eqref{eq_ebh}, we have the following terms: $J$ the tunnelling, $U$ the on-site two-atom interaction, $\epsilon$ the site energy, $\mu$ the chemical potential, and $V$ the two-site two-atom interaction. We will, unless otherwise stated, work in units of energy of the recoil energy $E_R = h^2/2m\lambda_{\mathrm{latt}}^2$, with $\lambda_{\mathrm{latt}}$ the optical lattice wavelength, and units of position of $\lambda_{\mathrm{latt}}$.

The properties of each lattice site in the tight-binding limit is given by the localised complete set of Wannier functions $w(\mathbf{r}-\mathbf{r}_i)$, where $\mathbf{r}-\mathbf{r}_i$ is the position away from the $i$th lattice site. The strength of each term in Hamiltonian~\eqref{eq_ebh} is dependent on an overlap integral between Wannier functions and the potential terms contained in Hamiltonian~\eqref{eq_fieldH}. The single-atom tunnelling strength is given by
\begin{equation}
\begin{aligned}
J_{ij} = & \, \dfrac{\hbar^2}{2m} \int d\mathbf{r}  w^*(\mathbf{r}-\mathbf{r}_i) \nabla^2 w(\mathbf{r} - \mathbf{r}_j) \\ & - \int d\mathbf{r}  w^*(\mathbf{r}-\mathbf{r}_i) V_{\mathrm{latt}}(\mathbf{r}) w(\mathbf{r}-\mathbf{r}_j),
\end{aligned}
\label{eq:Js}
\end{equation}
and the two-atom on-site interaction is
\begin{equation}
\begin{aligned}
U_i = & \, g \int d\mathbf{r} |w(\mathbf{r}-\mathbf{r}_i)|^4  \\ &+ \gamma \int d\mathbf{r}d\mathbf{r}' |w(\mathbf{r}-\mathbf{r}_i)|^2 h(\mathbf{r},\mathbf{r}') |w(\mathbf{r}'-\mathbf{r}_i)|^2.
\end{aligned}
\end{equation}
Due to the quasicrystalline and harmonic potential there is a position dependent site energy of
\begin{equation}
\begin{aligned}
\epsilon_i =  \int d \mathbf{r} |w(\mathbf{r} - \mathbf{r}_i)|^2 V_{\mathrm{latt}}(\mathbf{r}) \approx V_{\mathrm{latt}}(\mathbf{r}_i) .
\end{aligned}
\end{equation}
The two-site two-atom interaction strength is given by
\begin{equation}
\begin{aligned}
& V_{ij} = \dfrac{\gamma}{2} \int d\mathbf{r} d\mathbf{r}^\prime \Big[ |w(\mathbf{r} - \mathbf{r}_i)|^2 |w(\mathbf{r}' - \mathbf{r}_j)|^2 h(\mathbf{r},\mathbf{r}') \\ & + w^*(\mathbf{r} - \mathbf{r}_i) w^*(\mathbf{r}' - \mathbf{r}_j) w(\mathbf{r} - \mathbf{r}_j) w(\mathbf{r}' - \mathbf{r}_i)  h(\mathbf{r},\mathbf{r}') \Big].
\end{aligned}
\label{eq:Vs}
\end{equation}

\begin{figure}[t]
	\centering
	\includegraphics[width=0.98\linewidth]{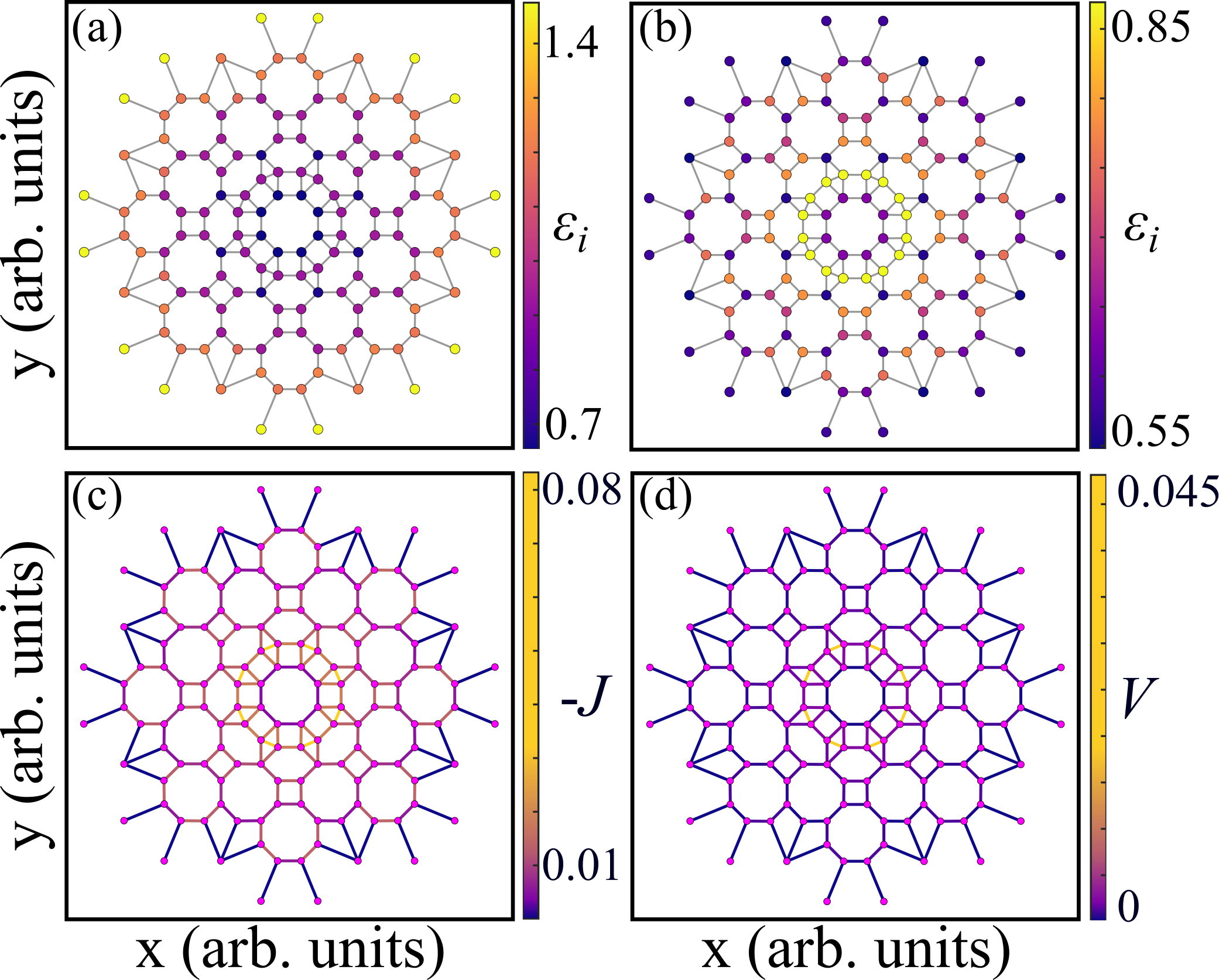}
	\caption{The strengths of terms in the extended Bose-Hubbard model for the eight-fold symmetric lattice. All energies are in units of $E_R$, and the lattice depth is $V_0 = 0.8E_R$. a) The energies $\epsilon_i$ due to the optical lattice plus harmonic potential. b) Same as (a) but excluding the harmonic trap. c) The site-dependent tunnelling strengths. d) The site-dependent long-range interaction in units of $\gamma E_R$.}
	\label{fig:NonHomogeneous}
\end{figure}

The position dependence of the parameter strengths of Hamiltonian~\eqref{eq_ebh} are shown in Fig.~\ref{fig:NonHomogeneous}. In these calculations we have taken a harmonic approximation for the individual wells of the lattice, with Wannier functions given by Gaussians. The harmonic approximation is known to provide good qualitative predictions, but will quantitatively underestimate certain parameters. However, as we will consider a Gutzwiller mean-field approach in this work it is already the case that only qualitative predictions should be expected. The site energy is dominated by the harmonic confinement, and even without the harmonic term the variation in the site energy is small, as shown in Figs.~\ref{fig:NonHomogeneous}a and b. Quasicrystalline potentials might lead to the observation of disorder induced phases like the Bose glass \cite{Fisher1989}. However, in the potential utilised for this work we will not observe any disorder induced ground state phases, which could be a result of the small differences in site energies across the lattice as shown in Fig.~\ref{fig:NonHomogeneous}a and b. 

In this work, we will assume that the physics of each individual site is dominated by its local neighbourhood. This should be a good assumption for the small lattice sizes we are considering in this work. We can then find the local nearest-neighbour couplings for each individual site by finding its nearest-neighbour which will be a distance $l$ away. We then couple the individual site bidirectionally to all sites that are of a distance $l$ apart. This results in the structure of bonds shown in Fig.~\ref{fig:NonHomogeneous}c, with each site coupled to its local nearest-neighbours. The tunnelling strengths shown in Fig.~\ref{fig:NonHomogeneous}c have a position dependence, which arises due to the incommensurate nature of the considered optical lattice. Another method of taking the tight-binding limit of a quasicrystal is to consider the tiles, and take all lines as bonds and vertices as lattice sites \cite{Repetowicz1998,Jagannathan2013}, which results in a similar lattice to that considered here.

\section{Gutzwiller mean-field}
\label{sec:Gutz}

The Gutzwiller mean-field is a well-known method to predict quantum phase transition in tight-binding models, and specifically Hubbard models \cite{Gutzwiller1963,Gutzwiller1964,Gutzwiller1965,Rokhsar1991,Krauth1992}. In the Gutzwiller mean-field approach, the many-body wave function is approximated as a product of on-site terms given by
\begin{equation}
\ket{\Psi} = \prod^{L}_i \sum^{z}_{n=0} f_n^{(i)} \ket{n_i},
\end{equation}
where $L$ defines the total number of lattice sites, $z$ is the maximum number of atoms per site, $f_n^{(i)}$ is the coefficient of having $n$ atoms in site $i$ and $\ket{n_i}$ is the corresponding Fock state. The coefficients are normalised such that $\sum^{z}_{n=0} |f_n^{(i)}|^2 = 1$. Correlations between sites are not accounted for due to the limiting of the Hilbert space \cite{lewenstein2012}. As a result, the Gutzwiller mean-field is known to give a good qualitative picture of phases in two dimensions, but the prediction of exact transition points is quantitatively unreliable.

To implement the Gutzwiller approach, we use a self-consistent loop which updates order parameters until convergence. This approach uses local order parameters, which are defined for each single site. The initialisation of the loop takes a random and uniformly distributed set of order parameters for each lattice site. We then diagonalise each local Hamiltonian with the given order parameters to solve for the local ground state. From this ground state, we update the local order parameters and then the loop continues to the next site, where the process is repeated. After one cycle through all lattice sites, we check for convergence towards the global ground state. If convergence is not achieved, the cycle through local lattice sites is repeated. Otherwise, the self-consistent loop ends.

\begin{figure}[t!]
	\centering
	\includegraphics[width=0.99\linewidth]{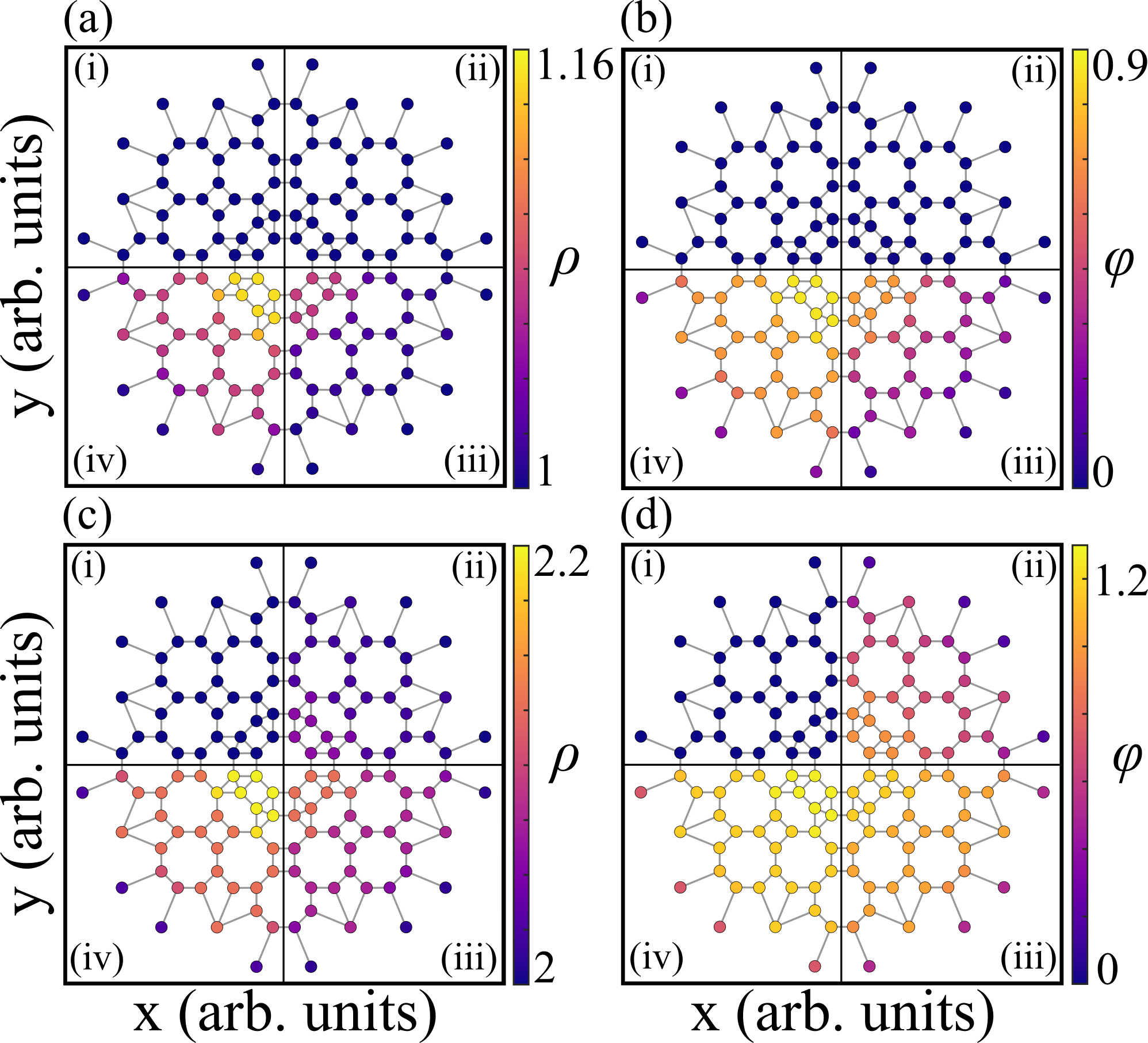}
	\caption{Mott-insulator to superfluid transition in the ground state of an eight-fold symmetric quasicrystal. (a,b) The case of $\mu/U = 0.5$ with (i) $J/U = 0.02$, (ii) $J/U = 0.04$, (iii) $J/U = 0.06$, and (iv) $J/U = 0.08$, showing an order one MI transitioning to a superfluid. (c,d) Same as the cases of (a,b) but for $\mu/U = 1.5$ and showing an order two MI transitioning to a superfluid at a different critical point. (a,c) Show the local density order parameters and (b,d) the local transport order parameter. Note, all ground states shown are eight-fold symmetric.}
	\label{fig:SFMI}
\end{figure}

\begin{figure}[t!]
	\centering
	\includegraphics[width=0.95\linewidth]{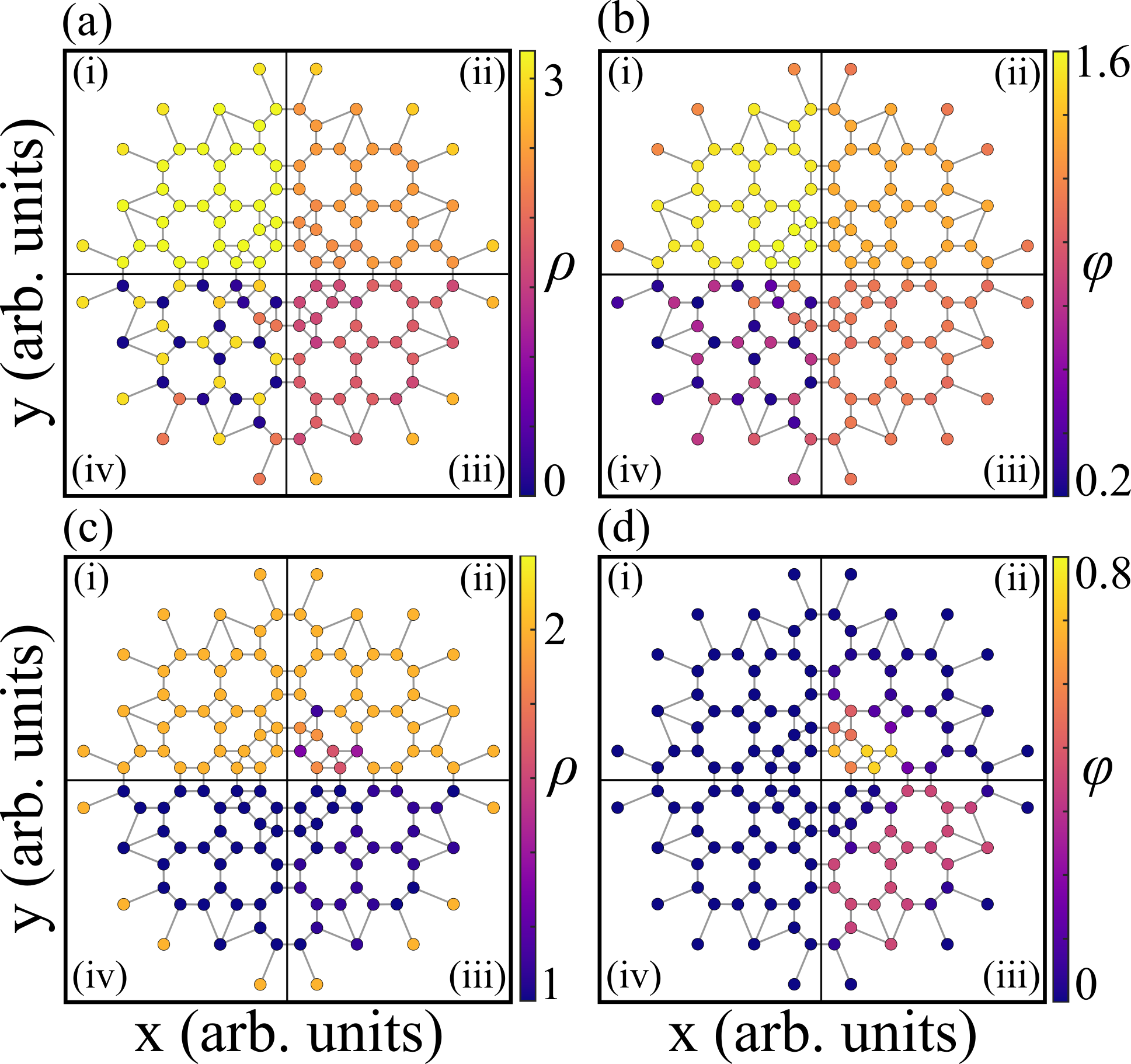}
	\caption{Phase transitions for non-zero long-range interactions in the ground state of an eight-fold symmetric quasicrystal. (a,b) The case of $\mu/U = 2.5$ and  $J/U = 0.1$ with (i) $V/U = 0$, (ii) $V/U = 0.1$, (iii) $V/U = 0.25$, and (iv) $V/U = 0.4$, showing a superfluid transitioning to a supersolid. (c,d) The case of $\mu/U = 1.5$ and  $J/U = 0.01$ with (i) $V/U = 0$, (ii) $V/U = 0.07$, (iii) $V/U = 0.17$, and (iv) $V/U = 0.25$, showing an order two Mott-insulator transitioning to a edge density wave. (a,c) Show the local density order parameters and (b,d) the local transport order parameter.}
	\label{fig:NonZeroV}
\end{figure}

\begin{figure}[t!]
	\centering
	\includegraphics[width=0.95\linewidth]{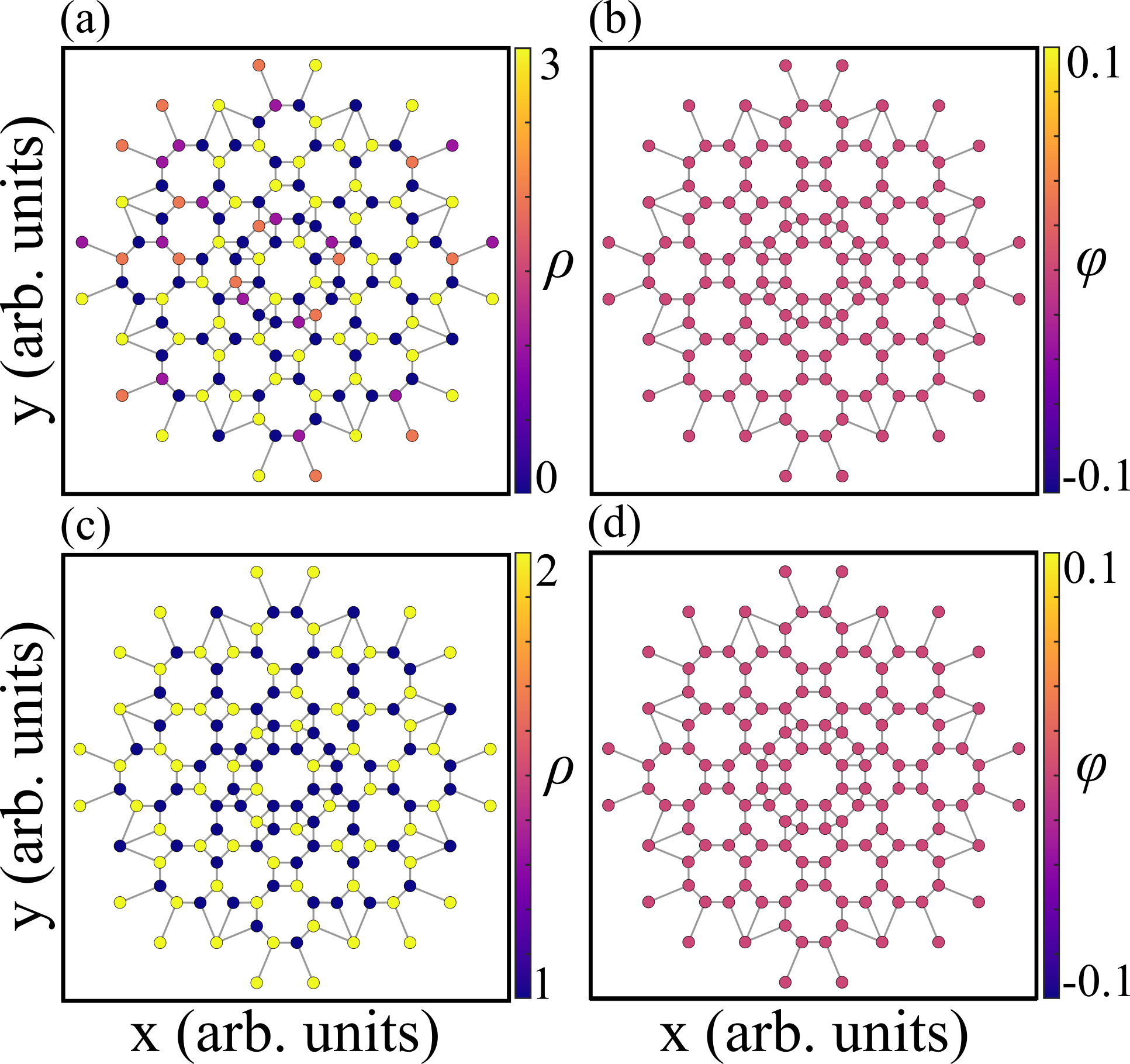}
	\caption{Density waves for non-zero long-range interactions in the ground state of an eight-fold symmetric quasicrystal. (a,b) The case of $\mu/U = 2.5$ and $J/U = 0.1$ with $V/U = 0.8$, showing a density wave of $\rho = 0$, $1$, $2$, and $3$ Mott states, note this ground state has no rotational symmetry. (c,d) The case of $\mu/U = 1.5$ and $J/U = 0.01$ with $V/U = 0.1$, showing a density wave of $\rho = 1$ and $2$ Mott states, note this ground state also has no rotational symmetry. (a,c) Show the local density order parameters and (b,d) the local transport order parameter.}
	\label{fig:DWs}
\end{figure}

For Hamiltonian~\eqref{eq_ebh}, there are two order parameters which govern the state of the system. These order parameters represent observables which can be measured in the system. The density $\rho_i$ is defined by
\begin{equation}
\rho_i = \langle \hat{n}_i \rangle = \sum^{z}_{n=0} n |f_n^{(i)}|^2,
\end{equation}
and the single atom transport $\varphi_i$ is defined by
\begin{equation}
\varphi_i = \langle \hat{b}_i \rangle = \sum^{z}_{n=0} \sqrt{n} f_n^{(i)} f_{n-1}^{*(i)}.
\end{equation}
In the extended Bose-Hubbard model, there are in general four phases that can be defined. The Mott-insulator with $\rho_i = \rho \: \forall i$ and $\varphi_i = 0 \: \forall i$, and the density wave with $\rho_i \neq \rho \: \forall i$ and $\varphi_i = 0 \: \forall i$, which both characterise phases with no transport properties. The superfluid with $\rho_i = \rho \: \forall i$ and $\varphi_i = \varphi \: \forall i$, and supersolid with $\rho_i \neq \rho \: \forall i$ and $\varphi_i \neq \varphi \: \forall i$, characterise ground states with non-zero transport properties. The definitions of the phases are the standard definitions considered for mean-field results \cite{Kimura2012,Kurdestany2012,Suzuki2014}. The supersolid and density wave are defined as usually having long-range crystalline order, which we will observe in the quasicrystal, with the ground state breaking the 8-fold symmetry in favour of a 4- or 2-fold symmetry when in the supersolid phase.

\section{Homogeneous system}

We will begin by taking the homogeneous case of strengths, with constant $J$, $U$, $V$ and $\epsilon$ throughout the lattice. While this is not a necessarily realistic scenario for when an ultracold gas is cooled into a quasicrystalline potential, it will provide a clear indication of the role that the local differences in the number of nearest neighbours plays in the phases. Note, we will also neglect the contribution of the harmonic potential to the on-site terms, which allows for the physics of the ground states due to the quasicrystalline potential to be investigated. While we have used the harmonic confinement to motivate the consideration of a finite system, it is worth noting that other confinment options, including box traps, are possible \cite{Gaunt2013,Nogrette2014}. These alternative traps could allow for finite systems without site energy contributions to be considered experimentally.

We will confirm that the phases discussed for the homogeneous case can be observed in the position-dependent Bose-Hubbard model in Sec.~\ref{sec:Real}. Of course, due to the local nature of the quasicrystal we can not consider only the usual $\mu-J$ plots of mean-field phase diagrams. Instead, we will visualise the local nature of the order parameters by plotting segments of the lattice, utilising the symmetries present to condense multiple phases onto single plots. Each plot, unless otherwise stated, contains four different parameter values plotted in segments of the quasicrystal, with the values labelled (i-iv) in an ascending order and plotted in a clockwise fashion. We will consider fixed chemical potentials on single lattice plots, with each of the four segments representing a varying of the tunnelling $J$ or long-range interaction $V$. In this section, we will work in the units of the contact interaction $U$ as we are treating the Bose-Hubbard model directly on the quasicrystal structure.

First, we consider the ubiquitous Mott-insulator to superfluid transition with $V=0$. This is shown in Fig.~\ref{fig:SFMI}, where we consider two values of $\mu$ ((a,b) $\mu/U=0.5$ and (c,d) $\mu/U = 1.5$) for a set of different $J$'s. The Mott-insulator to superfluid transition is observed clearly, and the transition is at different points for different values of $\mu$, as would be expected. The single-atom transport and density of the superfluid is not homogeneous in the lattice, and is therefore more generally a supersolid. This is a finite size effect with the modulation being small, and this state is therefore effectively a superfluid but on a finite size system \cite{Trefzger2011}. We observe the transition from Mott-insulator to superfluid effectively `spreading' through the system, as each site has in essence a different Mott to superfluid transition point due to the varying number of nearest neighbours. In this quasicystalline lattice, the superfluid is seen to initially populate the central portion of the lattice and then spreads over the lattice for increasing $J$. All ground states observed in Fig.~\ref{fig:SFMI} are eight-fold symmetric, preserving the quasicrystal symmetry. 

We then consider the case of non-zero two-site two-atom interactions in Fig.~\ref{fig:NonZeroV}. With a strong chemical potential (Fig.~\ref{fig:NonZeroV}a and b), we start with a standard superfluid (with finite size variations) for $V/U=0$. The non-local interactions make the bosonic atoms want to increase their separation, and with intermediate $V$ we observe a supersolid with a large density modulation favouring the edge of the system, see Fig.~\ref{fig:NonZeroV}a(iii). For large $V$, we observe an interesting supersolid phase which has a mix of three different local integer densities of $\rho = 0,2,3$. In this phase, the single-atom transport is non-zero for sites with non-zero density, and the state has spontaneously broken the lattice symmetry to be only four-fold symmetric, which is reminiscent of the staggered or twisted phases observed in standard lattice geometries \cite{Luhmann2016}. Taking the large $V$ limit, this phase retains its density structure and tends to a transport order parameter that is zero throughout the lattice. 

In Fig.~\ref{fig:NonZeroV}c and d, we consider an initial phase of a Mott-insulator then ramp up the two-site interaction strength. In this case, we observe a supersolid for intermediate $V$, see Fig.~\ref{fig:NonZeroV}c(ii), which is again four-fold symmetric. When the two-site interaction is increased a regime is reached where there is a $\rho = 1,2$ density wave on the edge of the lattice and the ground state is again eight-fold symmetric. This density wave is a combination of an order two and one Mott-insulator, with the order two on the edge to reduce the energetic cost of multiple atoms in neighbouring sites. Therefore, the edge density wave is a consequence of the varying number of nearest neighbour lattice sites in the quasicrystal and the long-range interactions.

We also observe intriguing intermediate density wave phases as shown in Fig.~\ref{fig:DWs}. These density waves are characterised by their uniform zero $\varphi$. In Fig.~\ref{fig:DWs}a we show an example where there is a combination of zero, first, second, and third order Mott-insulator state, whereas in Fig.~\ref{fig:DWs}c we observe a density wave of first and second order Mott states. We note that these density waves have no rotational symmetry, having spontaneously broken all rotational symmetries of the lattice. As these states do not possess any rotational or inversion symmetry, they do not fall into the categories of the usual crystal or quasicrystal phases, i.e. the order has been destroyed. The physics and dynamics of atoms in these phases with fully broken symmetry could be an interesting area of further study.

As mentioned at the start of this section, the homogeneous case allows for the consideration of the influence of the local variation in the number of nearest neighbours. Without non-local interactions, the variation in the number of neighbours results in slightly different local transition points of the Mott-insulator to superfluid. As can be seen in Fig.~\ref{fig:SFMI}, the superfluid state starts in the central portion of the lattice and propagates out when the tunnelling is increased. For non-zero non-local interactions, the number of nearest neighbours plays a more prominent role. This is especially evident in the highly structured density waves for strong non-local interactions observed in Fig.~\ref{fig:DWs}. In the symmetry breaking density waves of Fig.~\ref{fig:DWs} it is the variation in the number of nearest neighbours that causes the full symmetry breaking for the ground state. This can be observed by considering the local symmetries of the density wave. For some individual lattice sites it can be observed that there is a local rotational symmetry, e.g. a staggering of order parameters. This local rotational symmetry is not present for all lattice sites though, due to the variation in the number of nearest neighbours. This variation means it is difficult to obtain a staggering of the order parameters and to respect any global rotational symmetry, resulting in the ground state containing no rotational symmetry.

\section{Phase diagrams}

\begin{figure}[t]
	\centering
	\includegraphics[width=0.99\linewidth]{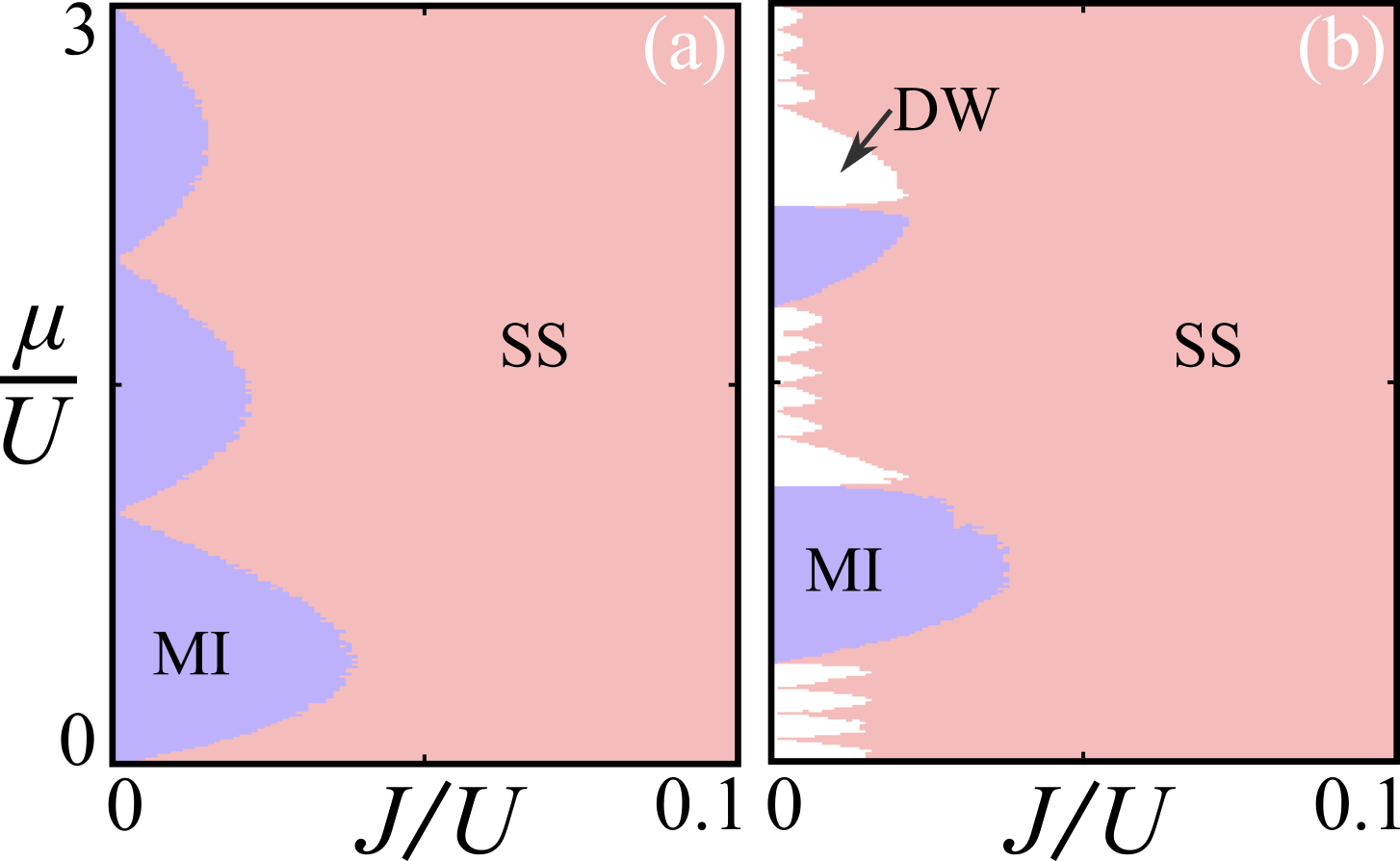}
	\caption{Phase diagrams for the homogeneous eight-fold symmetric quasicrystal. (a) The standard Bose-Hubbard model, exhibiting the expected MI and supersolid (SS) phases with the SS phase induced by the lattice geometry. (b) The extended Bose-Hubbard model with $V/U = 0.1$, exhibiting the expected MI, SS, and density wave (DW) phases, though the density wave phase takes a comb-like form in destroying the MI.}
	\label{fig:PhaseDiag}
\end{figure}

\begin{figure}[t]
	\centering
	\includegraphics[width=0.9\linewidth]{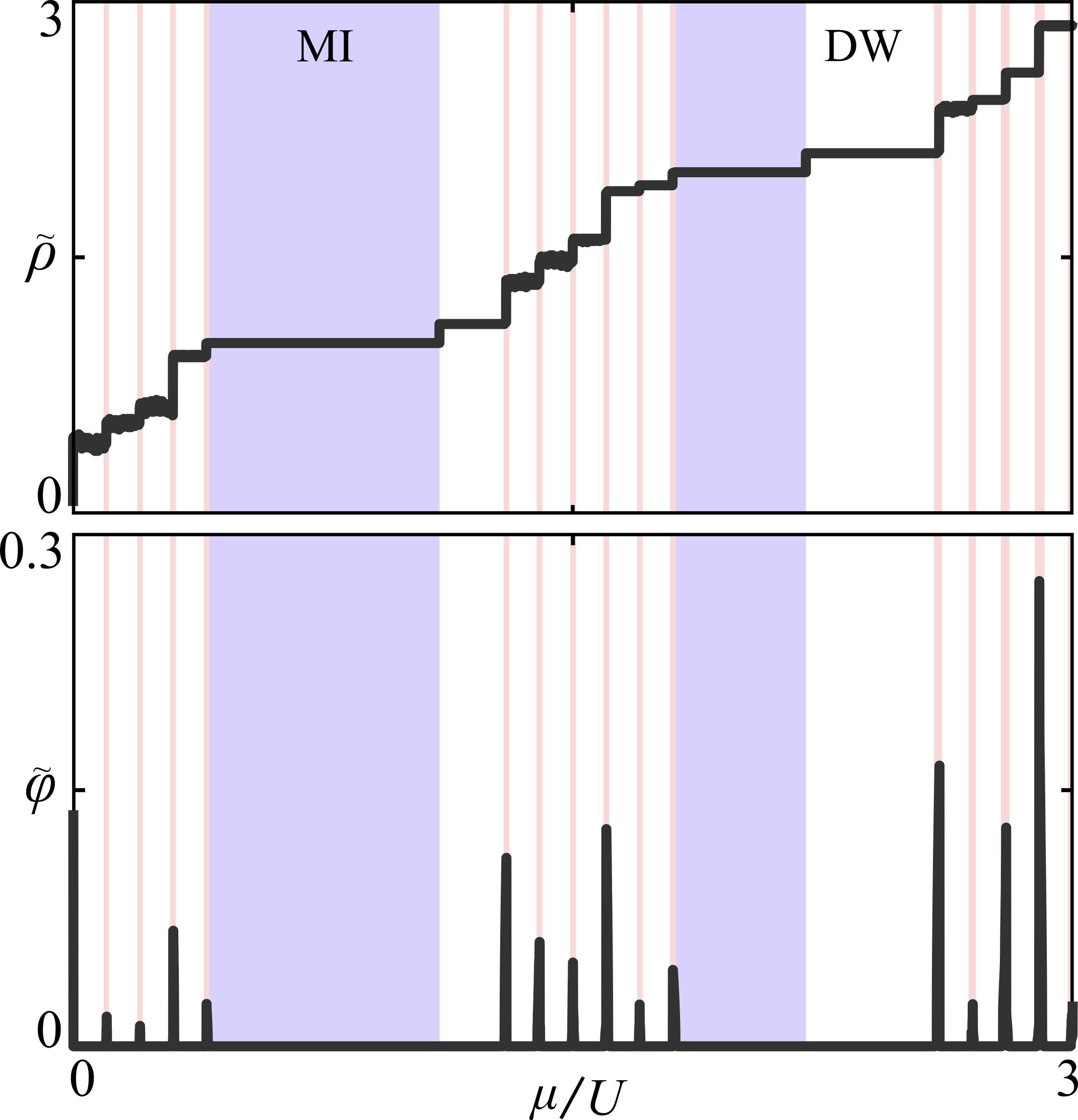}
	\caption{Plots of the average order parameters across the lattice for the homogeneous eight-fold symmetric quasicrystal. Considering $J/U = 0.001$ and $V/U=0.1$. (a) The average density $\tilde{\rho}$ across the lattice. (b) The average transport $\tilde{\varphi}$ across the lattice. Shaded regions correspond to DW, SS, and MI phases with the colours corresponding to those shown in Fig.~\ref{fig:PhaseDiag}b.}
	\label{fig:AverageCutOut}
\end{figure}

\begin{figure}[t]
	\centering
	\includegraphics[width=0.9\linewidth]{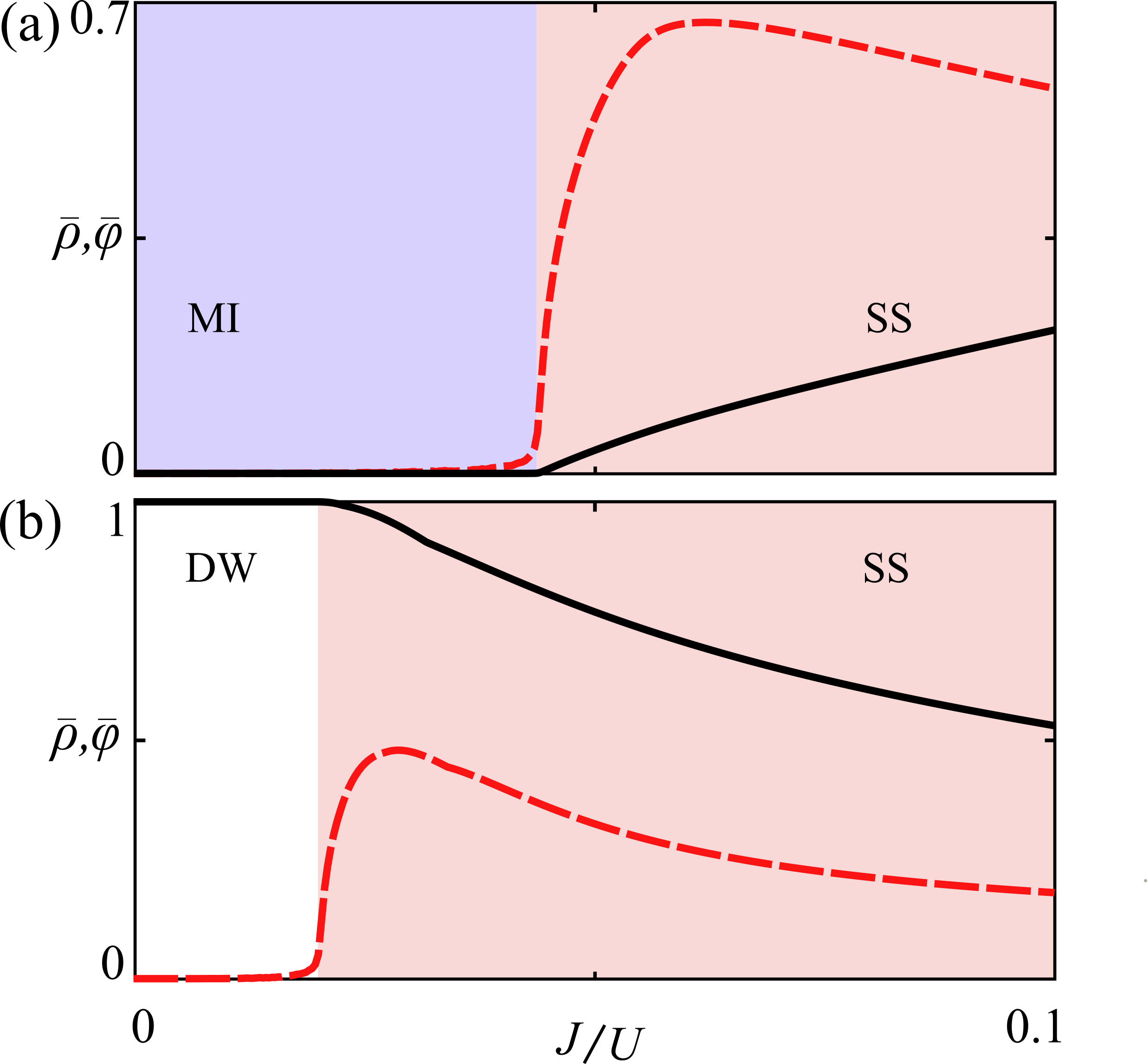}
	\caption{Plots of the difference of the order parameters across the lattice for the homogeneous eight-fold symmetric quasicrystal. The solid (black) line is $\bar{\rho}$, the difference between the maximum and mimimum value of $\rho$ across the lattice and dashed (red) line is $\bar{\varphi}$, the difference between the maximum and mimimum value of $\varphi$ across the lattice. (a) A cut along $\mu/U = 0.5$ of Fig.~\ref{fig:PhaseDiag}a. (b) A cut along $\mu/U = 2.4$ of Fig.~\ref{fig:PhaseDiag}b. The two cut-outs show clearly the existence of a single critical point.}
	\label{fig:PhaseCutOut}
\end{figure}

So far, we have considered a couple of different phase transitions and individual phases for specific values of the tunnelling and long-range interactions. It is usual for the consideration of phases in ultracold gases in Bose-Hubbard models to be plotted on $\mu-J$ diagrams. For the case of quasicrystals, these phase diagrams will not, as we have seen, tell the full story of the rotational symmetries present in the ground state. However, they will define the regions where the ground state belongs to one of the four possible phases. It is also known that the symmetry breaking phases discussed in the previous section will only occur in regions where the ground state is of the supersolid or density wave type. Therefore, knowing the regions where the supersolid and density wave phases dominate is vital to observe the intriguing symmetry breaking phases shown in Figs.~\ref{fig:NonZeroV} and \ref{fig:DWs}.

The phase diagrams are found by calculating the order parameters for a range of chemical potential and tunnelling. We then check these order parameters against a set of threshold conditions for the various phases as defined in Sec.~\ref{sec:Gutz}. The precise point of the phase boundaries is set by our threshold conditions and we assume that an order parameter is constant if it has variations that are less than $0.5\%$ of the maximum value of the order parameter for that ground state. We also require that there exists a critical point between each phase, and this rules out small SF regions that would be predicted by the threshold conditions but which are not physical.

In Fig.~\ref{fig:PhaseDiag}, we show two example phase diagrams of bosons in a quasicrystalline lattice for the homogeneous case, i.e. constant $J$ and $V$ with zero $\epsilon$. For the case of the usual Mott-insulator to superfluid transition, in Fig.~\ref{fig:PhaseDiag}a, we actually observe a Mott-insulator to supersolid transition. This supersolid phase is a direct result of the finite size of the quasicrystalline lattice, with the finite geometry favouring a supersolid over a superfluid phase. For the supersolid, the phase in the central portion of the lattice is a standard superfluid, with constant density, as is shown in Fig.~\ref{fig:SFMI}a(iv) and c(iv).

We also observe, in Fig.~\ref{fig:PhaseDiag}b, for non-zero $V$ that density waves destroy the standard Mott-insulating lobes as expected. For the case of non-zero $V$, the domination of the supersolid over the superfluid phases is already expected in regular lattice structures. In the quasicrystalline lattice, the supersolid is not only introduced by the two-site interactions but also by the aperiodic nature of the lattice. However, the density waves do not appear as single lobes replacing each of the Mott-insulator lobes as would be usually expected, but as a comb-like set of lobes. As usual, each density wave consists of the mixing of multiple order Mott-like states. The break-up of the usual lobe for each order of density wave is due to the quasicrystalline nature of the lattice. That is, discrete step-wise increases in the average density across the lattice are favoured due to the quasicrystalline lattice. To show this, we consider the average density, $\tilde{\rho}$, and transport, $\tilde{\varphi}$, across the lattice for a single tunnelling strength in Fig.~\ref{fig:AverageCutOut}. These plots show clearly the Mott-insulator (integer $\tilde{\rho}$), density wave (non-integer $\tilde{\rho}$), and supersolid (non-zero $\tilde{\varphi}$) phases. It is observed that each density wave lobe corresponds to a step-wise increase in the average density. This increase on the average corresponds to a higher proportion of the lattice being filled with higher $\rho$ Mott-like states and is a consequence of the varying number of nearest neighbours.

In Fig.~\ref{fig:PhaseCutOut}, we show two cut-outs of the variance in the order parameters for constant $\mu/U$. The variances are given by $\bar{\rho} = \max(\rho) - \min(\rho)$ and $\bar{\varphi} = \max(\varphi) - \min(\varphi)$. These cut-outs show the clear transition point between the phases.

\begin{figure}[t]
	\centering
	\includegraphics[width=0.99\linewidth]{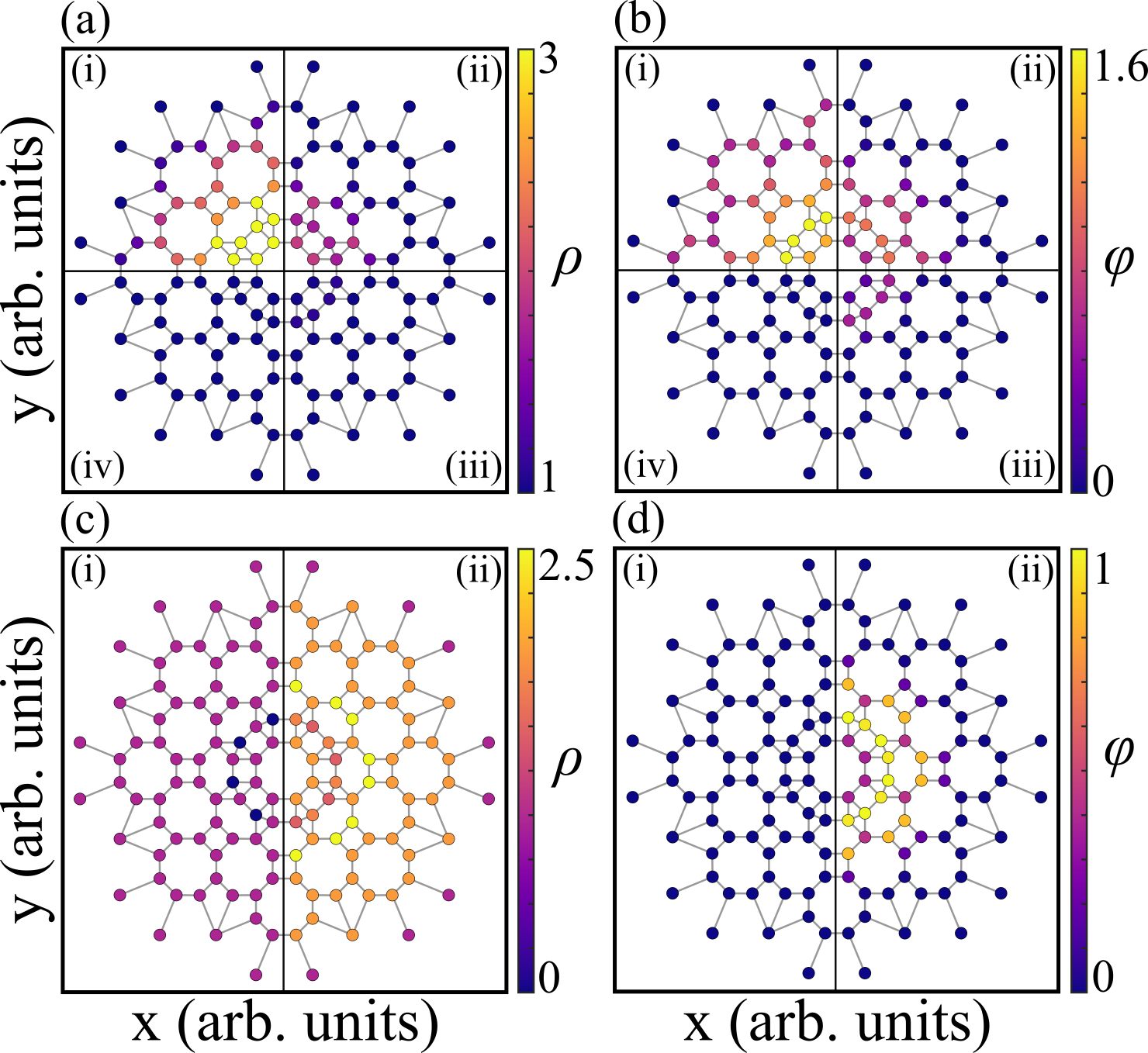}
	\caption{Phases for the position dependent eight-fold symmetric quasicrystal with non-homogeneous tunnelling, on-site and long-range interactions as shown in Fig.~\ref{fig:NonHomogeneous}. Lattice depth is set to $V_0=0.8 E_R$. (a,b) The case of $\mu = 1.5E_R$ with $\gamma=0$ and (i) $g = 5 E_R \lambda_{\mathrm{latt}}^3$, (ii) $g = 10 E_R \lambda_{\mathrm{latt}}^3$, (iii) $g = 15 E_R \lambda_{\mathrm{latt}}^3$, and (iv) $g = 20 E_R \lambda_{\mathrm{latt}}^3$, showing a superfluid transitioning to an order one MI. (c,d) The case of $\mu = 2.5E_R$ for (i) attractive $g = -2481 E_R \lambda_{\mathrm{latt}}^3$, $\gamma = 35E_R \lambda_{\mathrm{latt}}^3$, and (ii) attractive $g = -8684 E_R \lambda_{\mathrm{latt}}^3$, $\gamma = 10E_R \lambda_{\mathrm{latt}}^3$ showing a density wave to supersolid transition. (a,c) Show the local density order parameters and (b,d) the local transport order parameter.}
	\label{fig:NonHom}
\end{figure}

\begin{figure}[t]
	\centering
	\includegraphics[width=0.99\linewidth]{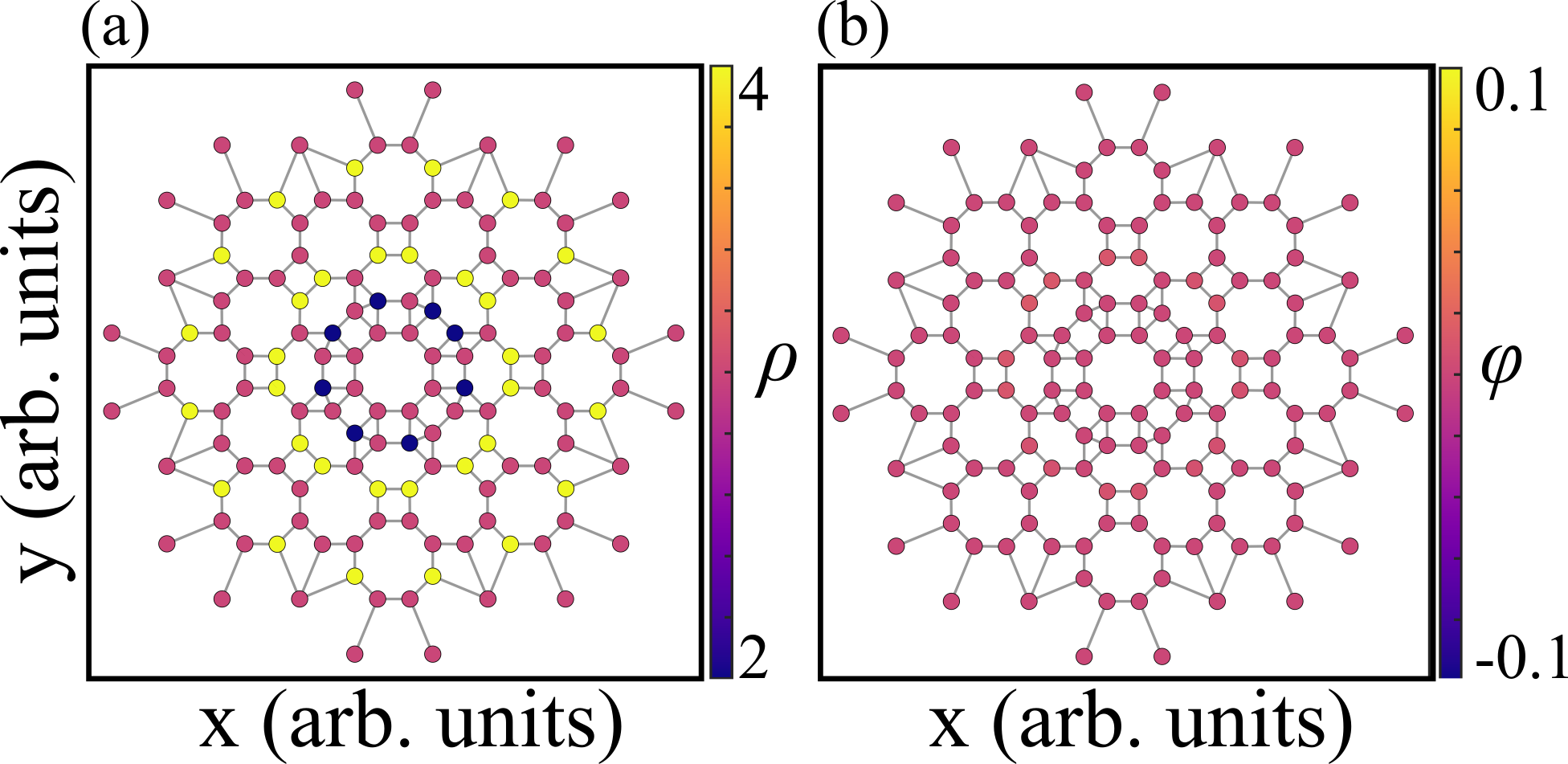}
	\caption{Example density wave for strong long range interactions for the position dependent eight-fold symmetric quasicrystal. We take the parameters $V_0=0.8E_R$, $\mu = 15E_R$, attractive $g = -12390 E_R \lambda_{\mathrm{latt}}^3$ and $\gamma=50E_R \lambda_{\mathrm{latt}}^3$. (a) Shows the local density order parameters and (b) the local transport order parameter.}
	\label{fig:DWsNonHom}
\end{figure}

\section{Position dependent system}
\label{sec:Real}

The case of an eight-fold optical lattice will naturally result in the various terms of the Bose-Hubbard model having position dependent strengths. This was discussed in Sec.~\ref{sec:Model} with the non-homogeneous strengths of site energy, tunnelling, and long range interactions shown in Fig.~\ref{fig:NonHomogeneous}. In this section, we consider if the phases of the previous sections can be observed when the position dependence of the strengths of the terms in the Bose-Hubbard model are taken into account. We have seen for the case of homogeneous strengths that various phases are possible in the extended Bose-Hubbard model and these phases can contain the eight-fold symmetry of the lattice, or spontaneously break the rotational symmetry. In this section, we will show that these phases can exist even when the non-homogeneous nature of the terms in the model are considered. Note, we are still considering the mean-field ground states, therefore, all quoted values are qualitative only and should be considered with care.

We will assume that the atoms sit in the ground states of the local minima of the quasicrystalline potential and take the harmonic approximation for each individual well. We will calculate the coefficients of the Bose-Hubbard models considered from Eqs.~\eqref{eq:Js} to~\eqref{eq:Vs}. Note, that this means that the constants set for the model of $g$ and $\gamma$ will now be in units of $E_R \lambda_{\mathrm{latt}}^3$. An important consequence of the non-uniform nature of the quasicrytalline lattice is that the ground state of each lattice site will be different, with a $0.3\lambda_{\mathrm{latt}}$ variation in the size of the harmonic states in our example lattice.

We will take a lattice depth of $V_0 = 0.8E_R$. Considering a chemical potential of $\mu = 1.5E_R$ and a two-body interaction of $g = 5 - 20 E_R \lambda_{\mathrm{latt}}^3$, in Fig.~\ref{fig:NonHom}a and b, we observe the standard superfluid to Mott-insulator transition. Note, the superfluid has a structure which is dominated by the harmonic trap but that still propagates out from the centre during the transition. By considering a fixed short range interaction and a variable long-range interaction of $\gamma = 35 - 10 E_R \lambda_{\mathrm{latt}}^3$ in Fig.~\ref{fig:NonHom}c and d, we can also observe a density wave to supersolid transition. The supersolid and density waves have densities which are not dominated by the harmonic trap, with the long-range interactions causing more interesting structures to appear. This includes a density wave in Fig.~\ref{fig:NonHom}c(i) which has spontaneously broken the eight-fold rotational symmetry of the lattice, and instead posses a mirror symmetry. The supersolid of Fig.~\ref{fig:NonHom}c(ii) is four-fold symmetric.

For large long-range interactions in the homogeneous case, we observed spontaneous symmetry breaking of all rotational symmetries of the quasicrystalline lattice. These interesting states can also be observed for the position-dependent case of non-homogeneous strengths with an example shown in Fig.~\ref{fig:DWsNonHom}. This ground state is a density wave with $\rho=2,3,4$ and possess no rotational symmetry due to an effective defect in the second inner-most ring. Higher order density waves can be observed for increased chemical potential. All examples including long-range interactions require a relative strength between the contact and long-range interactions of $\gamma/g \sim 10^{-3}$ which we would expect to be achievable with current techniques with dipolar atoms \cite{Trefzger2011,Menotti2008}. The small ratio required for the density wave phases is in part due to the requirement that the on-site contact and dipolar interactions cancel each other out.

\section{Conclusion}

With the recent realisation of a quasicrystalline optical lattice for ultracold atomic gases \cite{Viebahn2019}, the experimental investigation of quasicrystalline two-dimensional Bose-Hubbard models is within reach. We consider a toy model for the ground state phases of bosonic cold atoms in a quasicrystalline lattice. For the standard Bose-Hubbard model we observe the expected Mott-insulator phase as well as a supersolid phase. This supersolid phase is a direct result of the finite size of the lattice. We also consider the case of an extended Bose-Hubbard model, containing interactions between atoms in adjacent sites. This allows for density waves to also be observed. Both the supersolid and density wave phases can have interesting local structures that break the underlying eight-fold symmetry of the quasicrystal. For strong long-range interactions, we observe interesting density waves that spontaneously break all rotational symmetry.

The various intriguing phases observed in this work are a direct result of the aperiodic nature of the underlying lattice, with the variation of the local number of nearest neighbours being key. We have confirmed that in the eight-fold lattice with position dependent site energy, tunnelling, and long-range interactions, the various phases discussed in this work can occur. Also, this work shows that structurally complex phases can be observed in quasicrystalline lattices without invoking disorder. However, a future interesting path of study could be to consider any potential relationship between the observed ground state phases of this work, including those without rotational symmetry, and the predicted phases of disordered systems \cite{Fisher1989,Fallani2007}.

\begin{acknowledgments}
The authors would like to acknowledge helpful discussions with Manuel Valiente and Ulrich Schneider. D.J. and C.W.D. acknowledge support from EPSRC CM-CDT Grant No. EP/L015110/1.
\end{acknowledgments}



\begin{thebibliography}{68}%
\makeatletter
\providecommand \@ifxundefined [1]{%
 \@ifx{#1\undefined}
}%
\providecommand \@ifnum [1]{%
 \ifnum #1\expandafter \@firstoftwo
 \else \expandafter \@secondoftwo
 \fi
}%
\providecommand \@ifx [1]{%
 \ifx #1\expandafter \@firstoftwo
 \else \expandafter \@secondoftwo
 \fi
}%
\providecommand \natexlab [1]{#1}%
\providecommand \enquote  [1]{``#1''}%
\providecommand \bibnamefont  [1]{#1}%
\providecommand \bibfnamefont [1]{#1}%
\providecommand \citenamefont [1]{#1}%
\providecommand \href@noop [0]{\@secondoftwo}%
\providecommand \href [0]{\begingroup \@sanitize@url \@href}%
\providecommand \@href[1]{\@@startlink{#1}\@@href}%
\providecommand \@@href[1]{\endgroup#1\@@endlink}%
\providecommand \@sanitize@url [0]{\catcode `\\12\catcode `\$12\catcode
  `\&12\catcode `\#12\catcode `\^12\catcode `\_12\catcode `\%12\relax}%
\providecommand \@@startlink[1]{}%
\providecommand \@@endlink[0]{}%
\providecommand \url  [0]{\begingroup\@sanitize@url \@url }%
\providecommand \@url [1]{\endgroup\@href {#1}{\urlprefix }}%
\providecommand \urlprefix  [0]{URL }%
\providecommand \Eprint [0]{\href }%
\providecommand \doibase [0]{http://dx.doi.org/}%
\providecommand \selectlanguage [0]{\@gobble}%
\providecommand \bibinfo  [0]{\@secondoftwo}%
\providecommand \bibfield  [0]{\@secondoftwo}%
\providecommand \translation [1]{[#1]}%
\providecommand \BibitemOpen [0]{}%
\providecommand \bibitemStop [0]{}%
\providecommand \bibitemNoStop [0]{.\EOS\space}%
\providecommand \EOS [0]{\spacefactor3000\relax}%
\providecommand \BibitemShut  [1]{\csname bibitem#1\endcsname}%
\let\auto@bib@innerbib\@empty
\bibitem [{\citenamefont {Shechtman}\ \emph {et~al.}(1984)\citenamefont
  {Shechtman}, \citenamefont {Blech}, \citenamefont {Gratias},\ and\
  \citenamefont {Cahn}}]{Shechtman1984}%
  \BibitemOpen
  \bibfield  {author} {\bibinfo {author} {\bibfnamefont {D.}~\bibnamefont
  {Shechtman}}, \bibinfo {author} {\bibfnamefont {I.}~\bibnamefont {Blech}},
  \bibinfo {author} {\bibfnamefont {D.}~\bibnamefont {Gratias}}, \ and\
  \bibinfo {author} {\bibfnamefont {J.~W.}\ \bibnamefont {Cahn}},\ }\href
  {\doibase 10.1103/PhysRevLett.53.1951} {\bibfield  {journal} {\bibinfo
  {journal} {Phys. Rev. Lett.}\ }\textbf {\bibinfo {volume} {53}},\ \bibinfo
  {pages} {1951} (\bibinfo {year} {1984})}\BibitemShut {NoStop}%
\bibitem [{\citenamefont {Levine}\ and\ \citenamefont
  {Steinhardt}(1984)}]{Levine1984}%
  \BibitemOpen
  \bibfield  {author} {\bibinfo {author} {\bibfnamefont {D.}~\bibnamefont
  {Levine}}\ and\ \bibinfo {author} {\bibfnamefont {P.~J.}\ \bibnamefont
  {Steinhardt}},\ }\href {\doibase 10.1103/PhysRevLett.53.2477} {\bibfield
  {journal} {\bibinfo  {journal} {Phys. Rev. Lett.}\ }\textbf {\bibinfo
  {volume} {53}},\ \bibinfo {pages} {2477} (\bibinfo {year}
  {1984})}\BibitemShut {NoStop}%
\bibitem [{\citenamefont {Steurer}(2018)}]{steurer2018}%
  \BibitemOpen
  \bibfield  {author} {\bibinfo {author} {\bibfnamefont {W.}~\bibnamefont
  {Steurer}},\ }\href {\doibase 10.1107/S2053273317016540} {\bibfield
  {journal} {\bibinfo  {journal} {Acta Crystallogr. A}\ }\textbf {\bibinfo
  {volume} {74}},\ \bibinfo {pages} {1} (\bibinfo {year} {2018})}\BibitemShut
  {NoStop}%
\bibitem [{\citenamefont {Lang}\ \emph {et~al.}(2012)\citenamefont {Lang},
  \citenamefont {Cai},\ and\ \citenamefont {Chen}}]{Lang2012}%
  \BibitemOpen
  \bibfield  {author} {\bibinfo {author} {\bibfnamefont {L.-J.}\ \bibnamefont
  {Lang}}, \bibinfo {author} {\bibfnamefont {X.}~\bibnamefont {Cai}}, \ and\
  \bibinfo {author} {\bibfnamefont {S.}~\bibnamefont {Chen}},\ }\href {\doibase
  10.1103/PhysRevLett.108.220401} {\bibfield  {journal} {\bibinfo  {journal}
  {Phys. Rev. Lett.}\ }\textbf {\bibinfo {volume} {108}},\ \bibinfo {pages}
  {220401} (\bibinfo {year} {2012})}\BibitemShut {NoStop}%
\bibitem [{\citenamefont {Kraus}\ \emph {et~al.}(2013)\citenamefont {Kraus},
  \citenamefont {Ringel},\ and\ \citenamefont {Zilberberg}}]{Kraus2013}%
  \BibitemOpen
  \bibfield  {author} {\bibinfo {author} {\bibfnamefont {Y.~E.}\ \bibnamefont
  {Kraus}}, \bibinfo {author} {\bibfnamefont {Z.}~\bibnamefont {Ringel}}, \
  and\ \bibinfo {author} {\bibfnamefont {O.}~\bibnamefont {Zilberberg}},\
  }\href {\doibase 10.1103/PhysRevLett.111.226401} {\bibfield  {journal}
  {\bibinfo  {journal} {Phys. Rev. Lett.}\ }\textbf {\bibinfo {volume} {111}},\
  \bibinfo {pages} {226401} (\bibinfo {year} {2013})}\BibitemShut {NoStop}%
\bibitem [{\citenamefont {Penrose}(1974)}]{penrose1974}%
  \BibitemOpen
  \bibfield  {author} {\bibinfo {author} {\bibfnamefont {R.}~\bibnamefont
  {Penrose}},\ }\href@noop {} {\bibfield  {journal} {\bibinfo  {journal} {Bull.
  Inst. Math. Appl.}\ }\textbf {\bibinfo {volume} {10}},\ \bibinfo {pages}
  {266} (\bibinfo {year} {1974})}\BibitemShut {NoStop}%
\bibitem [{\citenamefont {Jenks}\ and\ \citenamefont
  {Thiel}(1998)}]{jenks1998}%
  \BibitemOpen
  \bibfield  {author} {\bibinfo {author} {\bibfnamefont {C.~J.}\ \bibnamefont
  {Jenks}}\ and\ \bibinfo {author} {\bibfnamefont {P.~A.}\ \bibnamefont
  {Thiel}},\ }\href {\doibase 10.1021/la970727+} {\bibfield  {journal}
  {\bibinfo  {journal} {Langmuir}\ }\textbf {\bibinfo {volume} {14}},\ \bibinfo
  {pages} {1392} (\bibinfo {year} {1998})}\BibitemShut {NoStop}%
\bibitem [{\citenamefont {Louzguine-Luzgin}\ and\ \citenamefont
  {Inoue}(2008)}]{Louzguine2008}%
  \BibitemOpen
  \bibfield  {author} {\bibinfo {author} {\bibfnamefont {D.}~\bibnamefont
  {Louzguine-Luzgin}}\ and\ \bibinfo {author} {\bibfnamefont {A.}~\bibnamefont
  {Inoue}},\ }\href {\doibase 10.1146/annurev.matsci.38.060407.130318}
  {\bibfield  {journal} {\bibinfo  {journal} {Ann. Rev. Mater. Res.}\ }\textbf
  {\bibinfo {volume} {38}},\ \bibinfo {pages} {403} (\bibinfo {year}
  {2008})}\BibitemShut {NoStop}%
\bibitem [{\citenamefont {Jin}\ and\ \citenamefont {Gao}(2012)}]{Jin2012}%
  \BibitemOpen
  \bibfield  {author} {\bibinfo {author} {\bibfnamefont {W.}~\bibnamefont
  {Jin}}\ and\ \bibinfo {author} {\bibfnamefont {Y.}~\bibnamefont {Gao}},\
  }\href {\doibase 10.1063/1.4754136} {\bibfield  {journal} {\bibinfo
  {journal} {Appl. Phys. Lett.}\ }\textbf {\bibinfo {volume} {101}},\ \bibinfo
  {pages} {141104} (\bibinfo {year} {2012})}\BibitemShut {NoStop}%
\bibitem [{\citenamefont {Vardeny}\ \emph {et~al.}(2013)\citenamefont
  {Vardeny}, \citenamefont {Nahata},\ and\ \citenamefont
  {Agrawal}}]{vardeny2013}%
  \BibitemOpen
  \bibfield  {author} {\bibinfo {author} {\bibfnamefont {Z.~V.}\ \bibnamefont
  {Vardeny}}, \bibinfo {author} {\bibfnamefont {A.}~\bibnamefont {Nahata}}, \
  and\ \bibinfo {author} {\bibfnamefont {A.}~\bibnamefont {Agrawal}},\ }\href
  {\doibase 10.1038/nphoton.2012.343} {\bibfield  {journal} {\bibinfo
  {journal} {Nat. Photonics}\ }\textbf {\bibinfo {volume} {7}},\ \bibinfo
  {pages} {177} (\bibinfo {year} {2013})}\BibitemShut {NoStop}%
\bibitem [{\citenamefont {Yao}\ \emph {et~al.}(2018)\citenamefont {Yao},
  \citenamefont {Wang}, \citenamefont {Bao}, \citenamefont {Zhang},
  \citenamefont {Zhang}, \citenamefont {Bao}, \citenamefont {Chan},
  \citenamefont {Chen}, \citenamefont {Avila}, \citenamefont {Asensio},
  \citenamefont {Zhu},\ and\ \citenamefont {Zhou}}]{Yao2018}%
  \BibitemOpen
  \bibfield  {author} {\bibinfo {author} {\bibfnamefont {W.}~\bibnamefont
  {Yao}}, \bibinfo {author} {\bibfnamefont {E.}~\bibnamefont {Wang}}, \bibinfo
  {author} {\bibfnamefont {C.}~\bibnamefont {Bao}}, \bibinfo {author}
  {\bibfnamefont {Y.}~\bibnamefont {Zhang}}, \bibinfo {author} {\bibfnamefont
  {K.}~\bibnamefont {Zhang}}, \bibinfo {author} {\bibfnamefont
  {K.}~\bibnamefont {Bao}}, \bibinfo {author} {\bibfnamefont {C.~K.}\
  \bibnamefont {Chan}}, \bibinfo {author} {\bibfnamefont {C.}~\bibnamefont
  {Chen}}, \bibinfo {author} {\bibfnamefont {J.}~\bibnamefont {Avila}},
  \bibinfo {author} {\bibfnamefont {M.~C.}\ \bibnamefont {Asensio}}, \bibinfo
  {author} {\bibfnamefont {J.}~\bibnamefont {Zhu}}, \ and\ \bibinfo {author}
  {\bibfnamefont {S.}~\bibnamefont {Zhou}},\ }\href {\doibase
  10.1073/pnas.1720865115} {\bibfield  {journal} {\bibinfo  {journal} {P. Natl.
  Acad. Sci.}\ }\textbf {\bibinfo {volume} {115}},\ \bibinfo {pages} {6928}
  (\bibinfo {year} {2018})}\BibitemShut {NoStop}%
\bibitem [{\citenamefont {Sakaguchi}\ and\ \citenamefont
  {Malomed}(2006)}]{Sakaguchi2006}%
  \BibitemOpen
  \bibfield  {author} {\bibinfo {author} {\bibfnamefont {H.}~\bibnamefont
  {Sakaguchi}}\ and\ \bibinfo {author} {\bibfnamefont {B.~A.}\ \bibnamefont
  {Malomed}},\ }\href {\doibase 10.1103/PhysRevE.74.026601} {\bibfield
  {journal} {\bibinfo  {journal} {Phys. Rev. E}\ }\textbf {\bibinfo {volume}
  {74}},\ \bibinfo {pages} {026601} (\bibinfo {year} {2006})}\BibitemShut
  {NoStop}%
\bibitem [{\citenamefont {Gopalakrishnan}\ \emph {et~al.}(2013)\citenamefont
  {Gopalakrishnan}, \citenamefont {Martin},\ and\ \citenamefont
  {Demler}}]{Gopalakrishnan2013}%
  \BibitemOpen
  \bibfield  {author} {\bibinfo {author} {\bibfnamefont {S.}~\bibnamefont
  {Gopalakrishnan}}, \bibinfo {author} {\bibfnamefont {I.}~\bibnamefont
  {Martin}}, \ and\ \bibinfo {author} {\bibfnamefont {E.~A.}\ \bibnamefont
  {Demler}},\ }\href {\doibase 10.1103/PhysRevLett.111.185304} {\bibfield
  {journal} {\bibinfo  {journal} {Phys. Rev. Lett.}\ }\textbf {\bibinfo
  {volume} {111}},\ \bibinfo {pages} {185304} (\bibinfo {year}
  {2013})}\BibitemShut {NoStop}%
\bibitem [{\citenamefont {Lu}\ \emph {et~al.}(1986)\citenamefont {Lu},
  \citenamefont {Odagaki},\ and\ \citenamefont {Birman}}]{Lu1986}%
  \BibitemOpen
  \bibfield  {author} {\bibinfo {author} {\bibfnamefont {J.~P.}\ \bibnamefont
  {Lu}}, \bibinfo {author} {\bibfnamefont {T.}~\bibnamefont {Odagaki}}, \ and\
  \bibinfo {author} {\bibfnamefont {J.~L.}\ \bibnamefont {Birman}},\ }\href
  {\doibase 10.1103/PhysRevB.33.4809} {\bibfield  {journal} {\bibinfo
  {journal} {Phys. Rev. B}\ }\textbf {\bibinfo {volume} {33}},\ \bibinfo
  {pages} {4809} (\bibinfo {year} {1986})}\BibitemShut {NoStop}%
\bibitem [{\citenamefont {Hiramoto}\ and\ \citenamefont
  {Kohmoto}(1992)}]{HIRAMOTO1992}%
  \BibitemOpen
  \bibfield  {author} {\bibinfo {author} {\bibfnamefont {H.}~\bibnamefont
  {Hiramoto}}\ and\ \bibinfo {author} {\bibfnamefont {M.}~\bibnamefont
  {Kohmoto}},\ }\href {\doibase 10.1142/S0217979292000153} {\bibfield
  {journal} {\bibinfo  {journal} {International Journal of Modern Physics B}\
  }\textbf {\bibinfo {volume} {06}},\ \bibinfo {pages} {281} (\bibinfo {year}
  {1992})}\BibitemShut {NoStop}%
\bibitem [{\citenamefont {Fallani}\ \emph {et~al.}(2007)\citenamefont
  {Fallani}, \citenamefont {Lye}, \citenamefont {Guarrera}, \citenamefont
  {Fort},\ and\ \citenamefont {Inguscio}}]{Fallani2007}%
  \BibitemOpen
  \bibfield  {author} {\bibinfo {author} {\bibfnamefont {L.}~\bibnamefont
  {Fallani}}, \bibinfo {author} {\bibfnamefont {J.~E.}\ \bibnamefont {Lye}},
  \bibinfo {author} {\bibfnamefont {V.}~\bibnamefont {Guarrera}}, \bibinfo
  {author} {\bibfnamefont {C.}~\bibnamefont {Fort}}, \ and\ \bibinfo {author}
  {\bibfnamefont {M.}~\bibnamefont {Inguscio}},\ }\href {\doibase
  10.1103/PhysRevLett.98.130404} {\bibfield  {journal} {\bibinfo  {journal}
  {Phys. Rev. Lett.}\ }\textbf {\bibinfo {volume} {98}},\ \bibinfo {pages}
  {130404} (\bibinfo {year} {2007})}\BibitemShut {NoStop}%
\bibitem [{\citenamefont {Roati}\ \emph {et~al.}(2008)\citenamefont {Roati},
  \citenamefont {D’Errico}, \citenamefont {Fallani}, \citenamefont {Fattori},
  \citenamefont {Fort}, \citenamefont {Zaccanti}, \citenamefont {Modugno},
  \citenamefont {Modugno},\ and\ \citenamefont {Inguscio}}]{roati2008}%
  \BibitemOpen
  \bibfield  {author} {\bibinfo {author} {\bibfnamefont {G.}~\bibnamefont
  {Roati}}, \bibinfo {author} {\bibfnamefont {C.}~\bibnamefont {D’Errico}},
  \bibinfo {author} {\bibfnamefont {L.}~\bibnamefont {Fallani}}, \bibinfo
  {author} {\bibfnamefont {M.}~\bibnamefont {Fattori}}, \bibinfo {author}
  {\bibfnamefont {C.}~\bibnamefont {Fort}}, \bibinfo {author} {\bibfnamefont
  {M.}~\bibnamefont {Zaccanti}}, \bibinfo {author} {\bibfnamefont
  {G.}~\bibnamefont {Modugno}}, \bibinfo {author} {\bibfnamefont
  {M.}~\bibnamefont {Modugno}}, \ and\ \bibinfo {author} {\bibfnamefont
  {M.}~\bibnamefont {Inguscio}},\ }\href {\doibase 10.1038/nature07071}
  {\bibfield  {journal} {\bibinfo  {journal} {Nature}\ }\textbf {\bibinfo
  {volume} {453}},\ \bibinfo {pages} {895} (\bibinfo {year}
  {2008})}\BibitemShut {NoStop}%
\bibitem [{\citenamefont {Edwards}\ \emph {et~al.}(2008)\citenamefont
  {Edwards}, \citenamefont {Beeler}, \citenamefont {Hong},\ and\ \citenamefont
  {Rolston}}]{Edwards2008}%
  \BibitemOpen
  \bibfield  {author} {\bibinfo {author} {\bibfnamefont {E.~E.}\ \bibnamefont
  {Edwards}}, \bibinfo {author} {\bibfnamefont {M.}~\bibnamefont {Beeler}},
  \bibinfo {author} {\bibfnamefont {T.}~\bibnamefont {Hong}}, \ and\ \bibinfo
  {author} {\bibfnamefont {S.~L.}\ \bibnamefont {Rolston}},\ }\href {\doibase
  10.1103/PhysRevLett.101.260402} {\bibfield  {journal} {\bibinfo  {journal}
  {Phys. Rev. Lett.}\ }\textbf {\bibinfo {volume} {101}},\ \bibinfo {pages}
  {260402} (\bibinfo {year} {2008})}\BibitemShut {NoStop}%
\bibitem [{\citenamefont {Gadway}\ \emph {et~al.}(2011)\citenamefont {Gadway},
  \citenamefont {Pertot}, \citenamefont {Reeves}, \citenamefont {Vogt},\ and\
  \citenamefont {Schneble}}]{Gadway2011}%
  \BibitemOpen
  \bibfield  {author} {\bibinfo {author} {\bibfnamefont {B.}~\bibnamefont
  {Gadway}}, \bibinfo {author} {\bibfnamefont {D.}~\bibnamefont {Pertot}},
  \bibinfo {author} {\bibfnamefont {J.}~\bibnamefont {Reeves}}, \bibinfo
  {author} {\bibfnamefont {M.}~\bibnamefont {Vogt}}, \ and\ \bibinfo {author}
  {\bibfnamefont {D.}~\bibnamefont {Schneble}},\ }\href {\doibase
  10.1103/PhysRevLett.107.145306} {\bibfield  {journal} {\bibinfo  {journal}
  {Phys. Rev. Lett.}\ }\textbf {\bibinfo {volume} {107}},\ \bibinfo {pages}
  {145306} (\bibinfo {year} {2011})}\BibitemShut {NoStop}%
\bibitem [{\citenamefont {D'Errico}\ \emph {et~al.}(2014)\citenamefont
  {D'Errico}, \citenamefont {Lucioni}, \citenamefont {Tanzi}, \citenamefont
  {Gori}, \citenamefont {Roux}, \citenamefont {McCulloch}, \citenamefont
  {Giamarchi}, \citenamefont {Inguscio},\ and\ \citenamefont
  {Modugno}}]{Errico2014}%
  \BibitemOpen
  \bibfield  {author} {\bibinfo {author} {\bibfnamefont {C.}~\bibnamefont
  {D'Errico}}, \bibinfo {author} {\bibfnamefont {E.}~\bibnamefont {Lucioni}},
  \bibinfo {author} {\bibfnamefont {L.}~\bibnamefont {Tanzi}}, \bibinfo
  {author} {\bibfnamefont {L.}~\bibnamefont {Gori}}, \bibinfo {author}
  {\bibfnamefont {G.}~\bibnamefont {Roux}}, \bibinfo {author} {\bibfnamefont
  {I.~P.}\ \bibnamefont {McCulloch}}, \bibinfo {author} {\bibfnamefont
  {T.}~\bibnamefont {Giamarchi}}, \bibinfo {author} {\bibfnamefont
  {M.}~\bibnamefont {Inguscio}}, \ and\ \bibinfo {author} {\bibfnamefont
  {G.}~\bibnamefont {Modugno}},\ }\href {\doibase
  10.1103/PhysRevLett.113.095301} {\bibfield  {journal} {\bibinfo  {journal}
  {Phys. Rev. Lett.}\ }\textbf {\bibinfo {volume} {113}},\ \bibinfo {pages}
  {095301} (\bibinfo {year} {2014})}\BibitemShut {NoStop}%
\bibitem [{\citenamefont {Singh}\ \emph {et~al.}(2015)\citenamefont {Singh},
  \citenamefont {Saha}, \citenamefont {Parameswaran},\ and\ \citenamefont
  {Weld}}]{Singh2015}%
  \BibitemOpen
  \bibfield  {author} {\bibinfo {author} {\bibfnamefont {K.}~\bibnamefont
  {Singh}}, \bibinfo {author} {\bibfnamefont {K.}~\bibnamefont {Saha}},
  \bibinfo {author} {\bibfnamefont {S.~A.}\ \bibnamefont {Parameswaran}}, \
  and\ \bibinfo {author} {\bibfnamefont {D.~M.}\ \bibnamefont {Weld}},\ }\href
  {\doibase 10.1103/PhysRevA.92.063426} {\bibfield  {journal} {\bibinfo
  {journal} {Phys. Rev. A}\ }\textbf {\bibinfo {volume} {92}},\ \bibinfo
  {pages} {063426} (\bibinfo {year} {2015})}\BibitemShut {NoStop}%
\bibitem [{\citenamefont {Schreiber}\ \emph {et~al.}(2015)\citenamefont
  {Schreiber}, \citenamefont {Hodgman}, \citenamefont {Bordia}, \citenamefont
  {L{\"u}schen}, \citenamefont {Fischer}, \citenamefont {Vosk}, \citenamefont
  {Altman}, \citenamefont {Schneider},\ and\ \citenamefont
  {Bloch}}]{schreiber2015}%
  \BibitemOpen
  \bibfield  {author} {\bibinfo {author} {\bibfnamefont {M.}~\bibnamefont
  {Schreiber}}, \bibinfo {author} {\bibfnamefont {S.~S.}\ \bibnamefont
  {Hodgman}}, \bibinfo {author} {\bibfnamefont {P.}~\bibnamefont {Bordia}},
  \bibinfo {author} {\bibfnamefont {H.~P.}\ \bibnamefont {L{\"u}schen}},
  \bibinfo {author} {\bibfnamefont {M.~H.}\ \bibnamefont {Fischer}}, \bibinfo
  {author} {\bibfnamefont {R.}~\bibnamefont {Vosk}}, \bibinfo {author}
  {\bibfnamefont {E.}~\bibnamefont {Altman}}, \bibinfo {author} {\bibfnamefont
  {U.}~\bibnamefont {Schneider}}, \ and\ \bibinfo {author} {\bibfnamefont
  {I.}~\bibnamefont {Bloch}},\ }\href {\doibase 10.1126/science.aaa7432}
  {\bibfield  {journal} {\bibinfo  {journal} {Science}\ }\textbf {\bibinfo
  {volume} {349}},\ \bibinfo {pages} {842} (\bibinfo {year}
  {2015})}\BibitemShut {NoStop}%
\bibitem [{\citenamefont {Hu}\ \emph {et~al.}(2016)\citenamefont {Hu},
  \citenamefont {Wang}, \citenamefont {Yi},\ and\ \citenamefont
  {Liu}}]{Hu2016}%
  \BibitemOpen
  \bibfield  {author} {\bibinfo {author} {\bibfnamefont {H.}~\bibnamefont
  {Hu}}, \bibinfo {author} {\bibfnamefont {A.-B.}\ \bibnamefont {Wang}},
  \bibinfo {author} {\bibfnamefont {S.}~\bibnamefont {Yi}}, \ and\ \bibinfo
  {author} {\bibfnamefont {X.-J.}\ \bibnamefont {Liu}},\ }\href {\doibase
  10.1103/PhysRevA.93.053601} {\bibfield  {journal} {\bibinfo  {journal} {Phys.
  Rev. A}\ }\textbf {\bibinfo {volume} {93}},\ \bibinfo {pages} {053601}
  (\bibinfo {year} {2016})}\BibitemShut {NoStop}%
\bibitem [{\citenamefont {Duncan}\ \emph {et~al.}(2017)\citenamefont {Duncan},
  \citenamefont {Loft}, \citenamefont {{\"O}hberg}, \citenamefont {Zinner},\
  and\ \citenamefont {Valiente}}]{Duncan2017}%
  \BibitemOpen
  \bibfield  {author} {\bibinfo {author} {\bibfnamefont {C.~W.}\ \bibnamefont
  {Duncan}}, \bibinfo {author} {\bibfnamefont {N.~J.~S.}\ \bibnamefont {Loft}},
  \bibinfo {author} {\bibfnamefont {P.}~\bibnamefont {{\"O}hberg}}, \bibinfo
  {author} {\bibfnamefont {N.~T.}\ \bibnamefont {Zinner}}, \ and\ \bibinfo
  {author} {\bibfnamefont {M.}~\bibnamefont {Valiente}},\ }\href {\doibase
  10.1007/s00601-016-1203-0} {\bibfield  {journal} {\bibinfo  {journal}
  {Few-Body Systems}\ }\textbf {\bibinfo {volume} {58}},\ \bibinfo {pages} {50}
  (\bibinfo {year} {2017})}\BibitemShut {NoStop}%
\bibitem [{\citenamefont {Valiente}\ \emph {et~al.}(2019)\citenamefont
  {Valiente}, \citenamefont {Duncan},\ and\ \citenamefont
  {Zinner}}]{valiente2019}%
  \BibitemOpen
  \bibfield  {author} {\bibinfo {author} {\bibfnamefont {M.}~\bibnamefont
  {Valiente}}, \bibinfo {author} {\bibfnamefont {C.~W.}\ \bibnamefont
  {Duncan}}, \ and\ \bibinfo {author} {\bibfnamefont {N.~T.}\ \bibnamefont
  {Zinner}},\ }\href@noop {} {\bibfield  {journal} {\bibinfo  {journal} {arXiv
  preprint arXiv:1908.03214}\ } (\bibinfo {year} {2019})}\BibitemShut {NoStop}%
\bibitem [{\citenamefont {Guidoni}\ \emph {et~al.}(1997)\citenamefont
  {Guidoni}, \citenamefont {Trich\'e}, \citenamefont {Verkerk},\ and\
  \citenamefont {Grynberg}}]{Guidoni1997}%
  \BibitemOpen
  \bibfield  {author} {\bibinfo {author} {\bibfnamefont {L.}~\bibnamefont
  {Guidoni}}, \bibinfo {author} {\bibfnamefont {C.}~\bibnamefont {Trich\'e}},
  \bibinfo {author} {\bibfnamefont {P.}~\bibnamefont {Verkerk}}, \ and\
  \bibinfo {author} {\bibfnamefont {G.}~\bibnamefont {Grynberg}},\ }\href
  {\doibase 10.1103/PhysRevLett.79.3363} {\bibfield  {journal} {\bibinfo
  {journal} {Phys. Rev. Lett.}\ }\textbf {\bibinfo {volume} {79}},\ \bibinfo
  {pages} {3363} (\bibinfo {year} {1997})}\BibitemShut {NoStop}%
\bibitem [{\citenamefont {Sanchez-Palencia}\ and\ \citenamefont
  {Santos}(2005)}]{Sanchez2005}%
  \BibitemOpen
  \bibfield  {author} {\bibinfo {author} {\bibfnamefont {L.}~\bibnamefont
  {Sanchez-Palencia}}\ and\ \bibinfo {author} {\bibfnamefont {L.}~\bibnamefont
  {Santos}},\ }\href {\doibase 10.1103/PhysRevA.72.053607} {\bibfield
  {journal} {\bibinfo  {journal} {Phys. Rev. A}\ }\textbf {\bibinfo {volume}
  {72}},\ \bibinfo {pages} {053607} (\bibinfo {year} {2005})}\BibitemShut
  {NoStop}%
\bibitem [{\citenamefont {Jagannathan}\ and\ \citenamefont
  {Duneau}(2013)}]{Jagannathan2013}%
  \BibitemOpen
  \bibfield  {author} {\bibinfo {author} {\bibfnamefont {A.}~\bibnamefont
  {Jagannathan}}\ and\ \bibinfo {author} {\bibfnamefont {M.}~\bibnamefont
  {Duneau}},\ }\href {\doibase 10.1209/0295-5075/104/66003} {\bibfield
  {journal} {\bibinfo  {journal} {Europhysics Lett. (EPL)}\ }\textbf {\bibinfo
  {volume} {104}},\ \bibinfo {pages} {66003} (\bibinfo {year}
  {2013})}\BibitemShut {NoStop}%
\bibitem [{\citenamefont {Hou}\ \emph {et~al.}(2018)\citenamefont {Hou},
  \citenamefont {Hu}, \citenamefont {Sun},\ and\ \citenamefont
  {Zhang}}]{Hou2018}%
  \BibitemOpen
  \bibfield  {author} {\bibinfo {author} {\bibfnamefont {J.}~\bibnamefont
  {Hou}}, \bibinfo {author} {\bibfnamefont {H.}~\bibnamefont {Hu}}, \bibinfo
  {author} {\bibfnamefont {K.}~\bibnamefont {Sun}}, \ and\ \bibinfo {author}
  {\bibfnamefont {C.}~\bibnamefont {Zhang}},\ }\href {\doibase
  10.1103/PhysRevLett.120.060407} {\bibfield  {journal} {\bibinfo  {journal}
  {Phys. Rev. Lett.}\ }\textbf {\bibinfo {volume} {120}},\ \bibinfo {pages}
  {060407} (\bibinfo {year} {2018})}\BibitemShut {NoStop}%
\bibitem [{\citenamefont {Corcovilos}\ and\ \citenamefont
  {Mittal}(2019)}]{Corcovilos2019}%
  \BibitemOpen
  \bibfield  {author} {\bibinfo {author} {\bibfnamefont {T.~A.}\ \bibnamefont
  {Corcovilos}}\ and\ \bibinfo {author} {\bibfnamefont {J.}~\bibnamefont
  {Mittal}},\ }\href {\doibase 10.1364/AO.58.002256} {\bibfield  {journal}
  {\bibinfo  {journal} {Appl. Opt.}\ }\textbf {\bibinfo {volume} {58}},\
  \bibinfo {pages} {2256} (\bibinfo {year} {2019})}\BibitemShut {NoStop}%
\bibitem [{\citenamefont {Viebahn}\ \emph {et~al.}(2019)\citenamefont
  {Viebahn}, \citenamefont {Sbroscia}, \citenamefont {Carter}, \citenamefont
  {Yu},\ and\ \citenamefont {Schneider}}]{Viebahn2019}%
  \BibitemOpen
  \bibfield  {author} {\bibinfo {author} {\bibfnamefont {K.}~\bibnamefont
  {Viebahn}}, \bibinfo {author} {\bibfnamefont {M.}~\bibnamefont {Sbroscia}},
  \bibinfo {author} {\bibfnamefont {E.}~\bibnamefont {Carter}}, \bibinfo
  {author} {\bibfnamefont {J.-C.}\ \bibnamefont {Yu}}, \ and\ \bibinfo {author}
  {\bibfnamefont {U.}~\bibnamefont {Schneider}},\ }\href {\doibase
  10.1103/PhysRevLett.122.110404} {\bibfield  {journal} {\bibinfo  {journal}
  {Phys. Rev. Lett.}\ }\textbf {\bibinfo {volume} {122}},\ \bibinfo {pages}
  {110404} (\bibinfo {year} {2019})}\BibitemShut {NoStop}%
\bibitem [{\citenamefont {Viebahn}(2018)}]{viebahn2018}%
  \BibitemOpen
  \bibfield  {author} {\bibinfo {author} {\bibfnamefont {K.~G.~H.}\
  \bibnamefont {Viebahn}},\ }\emph {\bibinfo {title} {Quasicrystalline optical
  lattices for ultracold atoms}},\ \href@noop {} {Ph.D. thesis},\ \bibinfo
  {school} {University of Cambridge} (\bibinfo {year} {2018})\BibitemShut
  {NoStop}%
\bibitem [{\citenamefont {Zhou}\ \emph {et~al.}(2018)\citenamefont {Zhou},
  \citenamefont {Jin},\ and\ \citenamefont {Schmiedmayer}}]{Zhou2018}%
  \BibitemOpen
  \bibfield  {author} {\bibinfo {author} {\bibfnamefont {X.}~\bibnamefont
  {Zhou}}, \bibinfo {author} {\bibfnamefont {S.}~\bibnamefont {Jin}}, \ and\
  \bibinfo {author} {\bibfnamefont {J.}~\bibnamefont {Schmiedmayer}},\ }\href
  {\doibase 10.1088/1367-2630/aac11b} {\bibfield  {journal} {\bibinfo
  {journal} {New J. Phys.}\ }\textbf {\bibinfo {volume} {20}},\ \bibinfo
  {pages} {055005} (\bibinfo {year} {2018})}\BibitemShut {NoStop}%
\bibitem [{\citenamefont {Khemani}\ \emph {et~al.}(2017)\citenamefont
  {Khemani}, \citenamefont {Sheng},\ and\ \citenamefont {Huse}}]{Khemani2017}%
  \BibitemOpen
  \bibfield  {author} {\bibinfo {author} {\bibfnamefont {V.}~\bibnamefont
  {Khemani}}, \bibinfo {author} {\bibfnamefont {D.~N.}\ \bibnamefont {Sheng}},
  \ and\ \bibinfo {author} {\bibfnamefont {D.~A.}\ \bibnamefont {Huse}},\
  }\href {\doibase 10.1103/PhysRevLett.119.075702} {\bibfield  {journal}
  {\bibinfo  {journal} {Phys. Rev. Lett.}\ }\textbf {\bibinfo {volume} {119}},\
  \bibinfo {pages} {075702} (\bibinfo {year} {2017})}\BibitemShut {NoStop}%
\bibitem [{\citenamefont {Fisher}\ \emph {et~al.}(1989)\citenamefont {Fisher},
  \citenamefont {Weichman}, \citenamefont {Grinstein},\ and\ \citenamefont
  {Fisher}}]{Fisher1989}%
  \BibitemOpen
  \bibfield  {author} {\bibinfo {author} {\bibfnamefont {M.~P.~A.}\
  \bibnamefont {Fisher}}, \bibinfo {author} {\bibfnamefont {P.~B.}\
  \bibnamefont {Weichman}}, \bibinfo {author} {\bibfnamefont {G.}~\bibnamefont
  {Grinstein}}, \ and\ \bibinfo {author} {\bibfnamefont {D.~S.}\ \bibnamefont
  {Fisher}},\ }\href {\doibase 10.1103/PhysRevB.40.546} {\bibfield  {journal}
  {\bibinfo  {journal} {Phys. Rev. B}\ }\textbf {\bibinfo {volume} {40}},\
  \bibinfo {pages} {546} (\bibinfo {year} {1989})}\BibitemShut {NoStop}%
\bibitem [{\citenamefont {Freericks}\ and\ \citenamefont
  {Monien}(1994)}]{Freericks1994}%
  \BibitemOpen
  \bibfield  {author} {\bibinfo {author} {\bibfnamefont {J.~K.}\ \bibnamefont
  {Freericks}}\ and\ \bibinfo {author} {\bibfnamefont {H.}~\bibnamefont
  {Monien}},\ }\href {\doibase 10.1209/0295-5075/26/7/012} {\bibfield
  {journal} {\bibinfo  {journal} {Europhys. Lett. ({EPL})}\ }\textbf {\bibinfo
  {volume} {26}},\ \bibinfo {pages} {545} (\bibinfo {year} {1994})}\BibitemShut
  {NoStop}%
\bibitem [{\citenamefont {Greiner}\ \emph {et~al.}(2002)\citenamefont
  {Greiner}, \citenamefont {Mandel}, \citenamefont {Esslinger}, \citenamefont
  {H{\"a}nsch},\ and\ \citenamefont {Bloch}}]{greiner2002}%
  \BibitemOpen
  \bibfield  {author} {\bibinfo {author} {\bibfnamefont {M.}~\bibnamefont
  {Greiner}}, \bibinfo {author} {\bibfnamefont {O.}~\bibnamefont {Mandel}},
  \bibinfo {author} {\bibfnamefont {T.}~\bibnamefont {Esslinger}}, \bibinfo
  {author} {\bibfnamefont {T.~W.}\ \bibnamefont {H{\"a}nsch}}, \ and\ \bibinfo
  {author} {\bibfnamefont {I.}~\bibnamefont {Bloch}},\ }\href {\doibase
  10.1038/415039a} {\bibfield  {journal} {\bibinfo  {journal} {Nature}\
  }\textbf {\bibinfo {volume} {415}},\ \bibinfo {pages} {39} (\bibinfo {year}
  {2002})}\BibitemShut {NoStop}%
\bibitem [{\citenamefont {Zwerger}(2003)}]{Zwerger2003}%
  \BibitemOpen
  \bibfield  {author} {\bibinfo {author} {\bibfnamefont {W.}~\bibnamefont
  {Zwerger}},\ }\href {\doibase 10.1088/1464-4266/5/2/352} {\bibfield
  {journal} {\bibinfo  {journal} {J. Opt. B Quantum S. O.}\ }\textbf {\bibinfo
  {volume} {5}},\ \bibinfo {pages} {S9} (\bibinfo {year} {2003})}\BibitemShut
  {NoStop}%
\bibitem [{\citenamefont {Bakr}\ \emph {et~al.}(2010)\citenamefont {Bakr},
  \citenamefont {Peng}, \citenamefont {Tai}, \citenamefont {Ma}, \citenamefont
  {Simon}, \citenamefont {Gillen}, \citenamefont {Foelling}, \citenamefont
  {Pollet},\ and\ \citenamefont {Greiner}}]{bakr2010}%
  \BibitemOpen
  \bibfield  {author} {\bibinfo {author} {\bibfnamefont {W.~S.}\ \bibnamefont
  {Bakr}}, \bibinfo {author} {\bibfnamefont {A.}~\bibnamefont {Peng}}, \bibinfo
  {author} {\bibfnamefont {M.~E.}\ \bibnamefont {Tai}}, \bibinfo {author}
  {\bibfnamefont {R.}~\bibnamefont {Ma}}, \bibinfo {author} {\bibfnamefont
  {J.}~\bibnamefont {Simon}}, \bibinfo {author} {\bibfnamefont {J.~I.}\
  \bibnamefont {Gillen}}, \bibinfo {author} {\bibfnamefont {S.}~\bibnamefont
  {Foelling}}, \bibinfo {author} {\bibfnamefont {L.}~\bibnamefont {Pollet}}, \
  and\ \bibinfo {author} {\bibfnamefont {M.}~\bibnamefont {Greiner}},\ }\href
  {\doibase 10.1126/science.1192368} {\bibfield  {journal} {\bibinfo  {journal}
  {Science}\ }\textbf {\bibinfo {volume} {329}},\ \bibinfo {pages} {547}
  (\bibinfo {year} {2010})}\BibitemShut {NoStop}%
\bibitem [{\citenamefont {Sowi\ifmmode~\acute{n}\else \'{n}\fi{}ski}\ \emph
  {et~al.}(2012)\citenamefont {Sowi\ifmmode~\acute{n}\else \'{n}\fi{}ski},
  \citenamefont {Dutta}, \citenamefont {Hauke}, \citenamefont {Tagliacozzo},\
  and\ \citenamefont {Lewenstein}}]{Tomasz2012}%
  \BibitemOpen
  \bibfield  {author} {\bibinfo {author} {\bibfnamefont {T.}~\bibnamefont
  {Sowi\ifmmode~\acute{n}\else \'{n}\fi{}ski}}, \bibinfo {author}
  {\bibfnamefont {O.}~\bibnamefont {Dutta}}, \bibinfo {author} {\bibfnamefont
  {P.}~\bibnamefont {Hauke}}, \bibinfo {author} {\bibfnamefont
  {L.}~\bibnamefont {Tagliacozzo}}, \ and\ \bibinfo {author} {\bibfnamefont
  {M.}~\bibnamefont {Lewenstein}},\ }\href {\doibase
  10.1103/PhysRevLett.108.115301} {\bibfield  {journal} {\bibinfo  {journal}
  {Phys. Rev. Lett.}\ }\textbf {\bibinfo {volume} {108}},\ \bibinfo {pages}
  {115301} (\bibinfo {year} {2012})}\BibitemShut {NoStop}%
\bibitem [{\citenamefont {Rossini}\ and\ \citenamefont
  {Fazio}(2012)}]{Rossini2012}%
  \BibitemOpen
  \bibfield  {author} {\bibinfo {author} {\bibfnamefont {D.}~\bibnamefont
  {Rossini}}\ and\ \bibinfo {author} {\bibfnamefont {R.}~\bibnamefont
  {Fazio}},\ }\href {\doibase 10.1088/1367-2630/14/6/065012} {\bibfield
  {journal} {\bibinfo  {journal} {New J. Phys.}\ }\textbf {\bibinfo {volume}
  {14}},\ \bibinfo {pages} {065012} (\bibinfo {year} {2012})}\BibitemShut
  {NoStop}%
\bibitem [{\citenamefont {Ohgoe}\ \emph {et~al.}(2012)\citenamefont {Ohgoe},
  \citenamefont {Suzuki},\ and\ \citenamefont {Kawashima}}]{Ohgoe2012}%
  \BibitemOpen
  \bibfield  {author} {\bibinfo {author} {\bibfnamefont {T.}~\bibnamefont
  {Ohgoe}}, \bibinfo {author} {\bibfnamefont {T.}~\bibnamefont {Suzuki}}, \
  and\ \bibinfo {author} {\bibfnamefont {N.}~\bibnamefont {Kawashima}},\ }\href
  {\doibase 10.1103/PhysRevB.86.054520} {\bibfield  {journal} {\bibinfo
  {journal} {Phys. Rev. B}\ }\textbf {\bibinfo {volume} {86}},\ \bibinfo
  {pages} {054520} (\bibinfo {year} {2012})}\BibitemShut {NoStop}%
\bibitem [{\citenamefont {Lewenstein}\ \emph {et~al.}(2012)\citenamefont
  {Lewenstein}, \citenamefont {Sanpera},\ and\ \citenamefont
  {Ahufinger}}]{lewenstein2012}%
  \BibitemOpen
  \bibfield  {author} {\bibinfo {author} {\bibfnamefont {M.}~\bibnamefont
  {Lewenstein}}, \bibinfo {author} {\bibfnamefont {A.}~\bibnamefont {Sanpera}},
  \ and\ \bibinfo {author} {\bibfnamefont {V.}~\bibnamefont {Ahufinger}},\
  }\href@noop {} {\emph {\bibinfo {title} {Ultracold Atoms in Optical Lattices:
  Simulating quantum many-body systems}}}\ (\bibinfo  {publisher} {Oxford
  University Press},\ \bibinfo {year} {2012})\ pp.\ \bibinfo {pages}
  {60--204}\BibitemShut {NoStop}%
\bibitem [{\citenamefont {Trefzger}\ \emph {et~al.}(2011)\citenamefont
  {Trefzger}, \citenamefont {Menotti}, \citenamefont {Capogrosso-Sansone},\
  and\ \citenamefont {Lewenstein}}]{Trefzger2011}%
  \BibitemOpen
  \bibfield  {author} {\bibinfo {author} {\bibfnamefont {C.}~\bibnamefont
  {Trefzger}}, \bibinfo {author} {\bibfnamefont {C.}~\bibnamefont {Menotti}},
  \bibinfo {author} {\bibfnamefont {B.}~\bibnamefont {Capogrosso-Sansone}}, \
  and\ \bibinfo {author} {\bibfnamefont {M.}~\bibnamefont {Lewenstein}},\
  }\href {\doibase 10.1088/0953-4075/44/19/193001} {\bibfield  {journal}
  {\bibinfo  {journal} {J. Phys. B At. Mol. Opt.}\ }\textbf {\bibinfo {volume}
  {44}},\ \bibinfo {pages} {193001} (\bibinfo {year} {2011})}\BibitemShut
  {NoStop}%
\bibitem [{\citenamefont {Aikawa}\ \emph {et~al.}(2012)\citenamefont {Aikawa},
  \citenamefont {Frisch}, \citenamefont {Mark}, \citenamefont {Baier},
  \citenamefont {Rietzler}, \citenamefont {Grimm},\ and\ \citenamefont
  {Ferlaino}}]{Aikawa2012}%
  \BibitemOpen
  \bibfield  {author} {\bibinfo {author} {\bibfnamefont {K.}~\bibnamefont
  {Aikawa}}, \bibinfo {author} {\bibfnamefont {A.}~\bibnamefont {Frisch}},
  \bibinfo {author} {\bibfnamefont {M.}~\bibnamefont {Mark}}, \bibinfo {author}
  {\bibfnamefont {S.}~\bibnamefont {Baier}}, \bibinfo {author} {\bibfnamefont
  {A.}~\bibnamefont {Rietzler}}, \bibinfo {author} {\bibfnamefont
  {R.}~\bibnamefont {Grimm}}, \ and\ \bibinfo {author} {\bibfnamefont
  {F.}~\bibnamefont {Ferlaino}},\ }\href {\doibase
  10.1103/PhysRevLett.108.210401} {\bibfield  {journal} {\bibinfo  {journal}
  {Phys. Rev. Lett.}\ }\textbf {\bibinfo {volume} {108}},\ \bibinfo {pages}
  {210401} (\bibinfo {year} {2012})}\BibitemShut {NoStop}%
\bibitem [{\citenamefont {Lu}\ \emph {et~al.}(2012)\citenamefont {Lu},
  \citenamefont {Burdick},\ and\ \citenamefont {Lev}}]{Lu2012}%
  \BibitemOpen
  \bibfield  {author} {\bibinfo {author} {\bibfnamefont {M.}~\bibnamefont
  {Lu}}, \bibinfo {author} {\bibfnamefont {N.~Q.}\ \bibnamefont {Burdick}}, \
  and\ \bibinfo {author} {\bibfnamefont {B.~L.}\ \bibnamefont {Lev}},\ }\href
  {\doibase 10.1103/PhysRevLett.108.215301} {\bibfield  {journal} {\bibinfo
  {journal} {Phys. Rev. Lett.}\ }\textbf {\bibinfo {volume} {108}},\ \bibinfo
  {pages} {215301} (\bibinfo {year} {2012})}\BibitemShut {NoStop}%
\bibitem [{\citenamefont {Baier}\ \emph {et~al.}(2016)\citenamefont {Baier},
  \citenamefont {Mark}, \citenamefont {Petter}, \citenamefont {Aikawa},
  \citenamefont {Chomaz}, \citenamefont {Cai}, \citenamefont {Baranov},
  \citenamefont {Zoller},\ and\ \citenamefont {Ferlaino}}]{baier2016}%
  \BibitemOpen
  \bibfield  {author} {\bibinfo {author} {\bibfnamefont {S.}~\bibnamefont
  {Baier}}, \bibinfo {author} {\bibfnamefont {M.~J.}\ \bibnamefont {Mark}},
  \bibinfo {author} {\bibfnamefont {D.}~\bibnamefont {Petter}}, \bibinfo
  {author} {\bibfnamefont {K.}~\bibnamefont {Aikawa}}, \bibinfo {author}
  {\bibfnamefont {L.}~\bibnamefont {Chomaz}}, \bibinfo {author} {\bibfnamefont
  {Z.}~\bibnamefont {Cai}}, \bibinfo {author} {\bibfnamefont {M.}~\bibnamefont
  {Baranov}}, \bibinfo {author} {\bibfnamefont {P.}~\bibnamefont {Zoller}}, \
  and\ \bibinfo {author} {\bibfnamefont {F.}~\bibnamefont {Ferlaino}},\ }\href
  {\doibase 10.1126/science.aac9812} {\bibfield  {journal} {\bibinfo  {journal}
  {Science}\ }\textbf {\bibinfo {volume} {352}},\ \bibinfo {pages} {201}
  (\bibinfo {year} {2016})}\BibitemShut {NoStop}%
\bibitem [{\citenamefont {Caballero-Benitez}\ and\ \citenamefont
  {Mekhov}(2016)}]{Caballero2016}%
  \BibitemOpen
  \bibfield  {author} {\bibinfo {author} {\bibfnamefont {S.~F.}\ \bibnamefont
  {Caballero-Benitez}}\ and\ \bibinfo {author} {\bibfnamefont {I.~B.}\
  \bibnamefont {Mekhov}},\ }\href {\doibase 10.1088/1367-2630/18/11/113010}
  {\bibfield  {journal} {\bibinfo  {journal} {New J. Phys.}\ }\textbf {\bibinfo
  {volume} {18}},\ \bibinfo {pages} {113010} (\bibinfo {year}
  {2016})}\BibitemShut {NoStop}%
\bibitem [{\citenamefont {Dogra}\ \emph {et~al.}(2016)\citenamefont {Dogra},
  \citenamefont {Brennecke}, \citenamefont {Huber},\ and\ \citenamefont
  {Donner}}]{Dogra2016}%
  \BibitemOpen
  \bibfield  {author} {\bibinfo {author} {\bibfnamefont {N.}~\bibnamefont
  {Dogra}}, \bibinfo {author} {\bibfnamefont {F.}~\bibnamefont {Brennecke}},
  \bibinfo {author} {\bibfnamefont {S.~D.}\ \bibnamefont {Huber}}, \ and\
  \bibinfo {author} {\bibfnamefont {T.}~\bibnamefont {Donner}},\ }\href
  {\doibase 10.1103/PhysRevA.94.023632} {\bibfield  {journal} {\bibinfo
  {journal} {Phys. Rev. A}\ }\textbf {\bibinfo {volume} {94}},\ \bibinfo
  {pages} {023632} (\bibinfo {year} {2016})}\BibitemShut {NoStop}%
\bibitem [{\citenamefont {Niederle}\ \emph {et~al.}(2016)\citenamefont
  {Niederle}, \citenamefont {Morigi},\ and\ \citenamefont
  {Rieger}}]{Niederle2016}%
  \BibitemOpen
  \bibfield  {author} {\bibinfo {author} {\bibfnamefont {A.~E.}\ \bibnamefont
  {Niederle}}, \bibinfo {author} {\bibfnamefont {G.}~\bibnamefont {Morigi}}, \
  and\ \bibinfo {author} {\bibfnamefont {H.}~\bibnamefont {Rieger}},\ }\href
  {\doibase 10.1103/PhysRevA.94.033607} {\bibfield  {journal} {\bibinfo
  {journal} {Phys. Rev. A}\ }\textbf {\bibinfo {volume} {94}},\ \bibinfo
  {pages} {033607} (\bibinfo {year} {2016})}\BibitemShut {NoStop}%
\bibitem [{\citenamefont {Landig}\ \emph {et~al.}(2016)\citenamefont {Landig},
  \citenamefont {Hruby}, \citenamefont {Dogra}, \citenamefont {Landini},
  \citenamefont {Mottl}, \citenamefont {Donner},\ and\ \citenamefont
  {Esslinger}}]{landig2016}%
  \BibitemOpen
  \bibfield  {author} {\bibinfo {author} {\bibfnamefont {R.}~\bibnamefont
  {Landig}}, \bibinfo {author} {\bibfnamefont {L.}~\bibnamefont {Hruby}},
  \bibinfo {author} {\bibfnamefont {N.}~\bibnamefont {Dogra}}, \bibinfo
  {author} {\bibfnamefont {M.}~\bibnamefont {Landini}}, \bibinfo {author}
  {\bibfnamefont {R.}~\bibnamefont {Mottl}}, \bibinfo {author} {\bibfnamefont
  {T.}~\bibnamefont {Donner}}, \ and\ \bibinfo {author} {\bibfnamefont
  {T.}~\bibnamefont {Esslinger}},\ }\href {\doibase 10.1038/nature17409}
  {\bibfield  {journal} {\bibinfo  {journal} {Nature}\ }\textbf {\bibinfo
  {volume} {532}},\ \bibinfo {pages} {476} (\bibinfo {year}
  {2016})}\BibitemShut {NoStop}%
\bibitem [{\citenamefont {Flottat}\ \emph {et~al.}(2017)\citenamefont
  {Flottat}, \citenamefont {de~Parny}, \citenamefont {H\'ebert}, \citenamefont
  {Rousseau},\ and\ \citenamefont {Batrouni}}]{Flottat2017}%
  \BibitemOpen
  \bibfield  {author} {\bibinfo {author} {\bibfnamefont {T.}~\bibnamefont
  {Flottat}}, \bibinfo {author} {\bibfnamefont {L.~d.~F.}\ \bibnamefont
  {de~Parny}}, \bibinfo {author} {\bibfnamefont {F.}~\bibnamefont {H\'ebert}},
  \bibinfo {author} {\bibfnamefont {V.~G.}\ \bibnamefont {Rousseau}}, \ and\
  \bibinfo {author} {\bibfnamefont {G.~G.}\ \bibnamefont {Batrouni}},\ }\href
  {\doibase 10.1103/PhysRevB.95.144501} {\bibfield  {journal} {\bibinfo
  {journal} {Phys. Rev. B}\ }\textbf {\bibinfo {volume} {95}},\ \bibinfo
  {pages} {144501} (\bibinfo {year} {2017})}\BibitemShut {NoStop}%
\bibitem [{\citenamefont {Landau}\ and\ \citenamefont
  {Lifshitz}(1958)}]{Landau1958}%
  \BibitemOpen
  \bibfield  {author} {\bibinfo {author} {\bibfnamefont {L.}~\bibnamefont
  {Landau}}\ and\ \bibinfo {author} {\bibfnamefont {E.}~\bibnamefont
  {Lifshitz}},\ }\href@noop {} {\emph {\bibinfo {title} {Statistical Physics
  (Course of Theoretical Physics vol 5)}}},\ \bibinfo {edition} {1st}\ ed.\
  (\bibinfo  {publisher} {Addison-Wesley publishing company},\ \bibinfo {year}
  {1958})\ pp.\ \bibinfo {pages} {409--429}\BibitemShut {NoStop}%
\bibitem [{\citenamefont {Jaksch}\ \emph {et~al.}(1998)\citenamefont {Jaksch},
  \citenamefont {Bruder}, \citenamefont {Cirac}, \citenamefont {Gardiner},\
  and\ \citenamefont {Zoller}}]{Jaksch1998}%
  \BibitemOpen
  \bibfield  {author} {\bibinfo {author} {\bibfnamefont {D.}~\bibnamefont
  {Jaksch}}, \bibinfo {author} {\bibfnamefont {C.}~\bibnamefont {Bruder}},
  \bibinfo {author} {\bibfnamefont {J.~I.}\ \bibnamefont {Cirac}}, \bibinfo
  {author} {\bibfnamefont {C.~W.}\ \bibnamefont {Gardiner}}, \ and\ \bibinfo
  {author} {\bibfnamefont {P.}~\bibnamefont {Zoller}},\ }\href {\doibase
  10.1103/PhysRevLett.81.3108} {\bibfield  {journal} {\bibinfo  {journal}
  {Phys. Rev. Lett.}\ }\textbf {\bibinfo {volume} {81}},\ \bibinfo {pages}
  {3108} (\bibinfo {year} {1998})}\BibitemShut {NoStop}%
\bibitem [{\citenamefont {Dutta}\ \emph {et~al.}(2015)\citenamefont {Dutta},
  \citenamefont {Gajda}, \citenamefont {Hauke}, \citenamefont {Lewenstein},
  \citenamefont {Lühmann}, \citenamefont {Malomed}, \citenamefont
  {Sowi{\'{n}}ski},\ and\ \citenamefont {Zakrzewski}}]{Dutta2015}%
  \BibitemOpen
  \bibfield  {author} {\bibinfo {author} {\bibfnamefont {O.}~\bibnamefont
  {Dutta}}, \bibinfo {author} {\bibfnamefont {M.}~\bibnamefont {Gajda}},
  \bibinfo {author} {\bibfnamefont {P.}~\bibnamefont {Hauke}}, \bibinfo
  {author} {\bibfnamefont {M.}~\bibnamefont {Lewenstein}}, \bibinfo {author}
  {\bibfnamefont {D.-S.}\ \bibnamefont {Lühmann}}, \bibinfo {author}
  {\bibfnamefont {B.~A.}\ \bibnamefont {Malomed}}, \bibinfo {author}
  {\bibfnamefont {T.}~\bibnamefont {Sowi{\'{n}}ski}}, \ and\ \bibinfo {author}
  {\bibfnamefont {J.}~\bibnamefont {Zakrzewski}},\ }\href {\doibase
  10.1088/0034-4885/78/6/066001} {\bibfield  {journal} {\bibinfo  {journal}
  {Rep. Prog. Phys.}\ }\textbf {\bibinfo {volume} {78}},\ \bibinfo {pages}
  {066001} (\bibinfo {year} {2015})}\BibitemShut {NoStop}%
\bibitem [{\citenamefont {Repetowicz}\ \emph {et~al.}(1998)\citenamefont
  {Repetowicz}, \citenamefont {Grimm},\ and\ \citenamefont
  {Schreiber}}]{Repetowicz1998}%
  \BibitemOpen
  \bibfield  {author} {\bibinfo {author} {\bibfnamefont {P.}~\bibnamefont
  {Repetowicz}}, \bibinfo {author} {\bibfnamefont {U.}~\bibnamefont {Grimm}}, \
  and\ \bibinfo {author} {\bibfnamefont {M.}~\bibnamefont {Schreiber}},\ }\href
  {\doibase 10.1103/PhysRevB.58.13482} {\bibfield  {journal} {\bibinfo
  {journal} {Phys. Rev. B}\ }\textbf {\bibinfo {volume} {58}},\ \bibinfo
  {pages} {13482} (\bibinfo {year} {1998})}\BibitemShut {NoStop}%
\bibitem [{\citenamefont {Gutzwiller}(1963)}]{Gutzwiller1963}%
  \BibitemOpen
  \bibfield  {author} {\bibinfo {author} {\bibfnamefont {M.~C.}\ \bibnamefont
  {Gutzwiller}},\ }\href {\doibase 10.1103/PhysRevLett.10.159} {\bibfield
  {journal} {\bibinfo  {journal} {Phys. Rev. Lett.}\ }\textbf {\bibinfo
  {volume} {10}},\ \bibinfo {pages} {159} (\bibinfo {year} {1963})}\BibitemShut
  {NoStop}%
\bibitem [{\citenamefont {Gutzwiller}(1964)}]{Gutzwiller1964}%
  \BibitemOpen
  \bibfield  {author} {\bibinfo {author} {\bibfnamefont {M.~C.}\ \bibnamefont
  {Gutzwiller}},\ }\href {\doibase 10.1103/PhysRev.134.A923} {\bibfield
  {journal} {\bibinfo  {journal} {Phys. Rev.}\ }\textbf {\bibinfo {volume}
  {134}},\ \bibinfo {pages} {A923} (\bibinfo {year} {1964})}\BibitemShut
  {NoStop}%
\bibitem [{\citenamefont {Gutzwiller}(1965)}]{Gutzwiller1965}%
  \BibitemOpen
  \bibfield  {author} {\bibinfo {author} {\bibfnamefont {M.~C.}\ \bibnamefont
  {Gutzwiller}},\ }\href {\doibase 10.1103/PhysRev.137.A1726} {\bibfield
  {journal} {\bibinfo  {journal} {Phys. Rev.}\ }\textbf {\bibinfo {volume}
  {137}},\ \bibinfo {pages} {A1726} (\bibinfo {year} {1965})}\BibitemShut
  {NoStop}%
\bibitem [{\citenamefont {Rokhsar}\ and\ \citenamefont
  {Kotliar}(1991)}]{Rokhsar1991}%
  \BibitemOpen
  \bibfield  {author} {\bibinfo {author} {\bibfnamefont {D.~S.}\ \bibnamefont
  {Rokhsar}}\ and\ \bibinfo {author} {\bibfnamefont {B.~G.}\ \bibnamefont
  {Kotliar}},\ }\href {\doibase 10.1103/PhysRevB.44.10328} {\bibfield
  {journal} {\bibinfo  {journal} {Phys. Rev. B}\ }\textbf {\bibinfo {volume}
  {44}},\ \bibinfo {pages} {10328} (\bibinfo {year} {1991})}\BibitemShut
  {NoStop}%
\bibitem [{\citenamefont {Krauth}\ \emph {et~al.}(1992)\citenamefont {Krauth},
  \citenamefont {Caffarel},\ and\ \citenamefont {Bouchaud}}]{Krauth1992}%
  \BibitemOpen
  \bibfield  {author} {\bibinfo {author} {\bibfnamefont {W.}~\bibnamefont
  {Krauth}}, \bibinfo {author} {\bibfnamefont {M.}~\bibnamefont {Caffarel}}, \
  and\ \bibinfo {author} {\bibfnamefont {J.-P.}\ \bibnamefont {Bouchaud}},\
  }\href {\doibase 10.1103/PhysRevB.45.3137} {\bibfield  {journal} {\bibinfo
  {journal} {Phys. Rev. B}\ }\textbf {\bibinfo {volume} {45}},\ \bibinfo
  {pages} {3137} (\bibinfo {year} {1992})}\BibitemShut {NoStop}%
\bibitem [{\citenamefont {Kimura}(2012)}]{Kimura2012}%
  \BibitemOpen
  \bibfield  {author} {\bibinfo {author} {\bibfnamefont {T.}~\bibnamefont
  {Kimura}},\ }\href {\doibase 10.1088/1742-6596/400/1/012032} {\bibfield
  {journal} {\bibinfo  {journal} {Journal of Physics: Conference Series}\
  }\textbf {\bibinfo {volume} {400}},\ \bibinfo {pages} {012032} (\bibinfo
  {year} {2012})}\BibitemShut {NoStop}%
\bibitem [{\citenamefont {Kurdestany}\ \emph {et~al.}(2012)\citenamefont
  {Kurdestany}, \citenamefont {Pai},\ and\ \citenamefont
  {Pandit}}]{Kurdestany2012}%
  \BibitemOpen
  \bibfield  {author} {\bibinfo {author} {\bibfnamefont {J.}~\bibnamefont
  {Kurdestany}}, \bibinfo {author} {\bibfnamefont {R.}~\bibnamefont {Pai}}, \
  and\ \bibinfo {author} {\bibfnamefont {R.}~\bibnamefont {Pandit}},\ }\href
  {\doibase 10.1002/andp.201100274} {\bibfield  {journal} {\bibinfo  {journal}
  {Annalen der Physik}\ }\textbf {\bibinfo {volume} {524}},\ \bibinfo {pages}
  {234} (\bibinfo {year} {2012})}\BibitemShut {NoStop}%
\bibitem [{\citenamefont {Suzuki}\ and\ \citenamefont
  {Koga}(2014)}]{Suzuki2014}%
  \BibitemOpen
  \bibfield  {author} {\bibinfo {author} {\bibfnamefont {R.}~\bibnamefont
  {Suzuki}}\ and\ \bibinfo {author} {\bibfnamefont {A.}~\bibnamefont {Koga}},\
  }\href {\doibase 10.7566/JPSJ.83.064003} {\bibfield  {journal} {\bibinfo
  {journal} {Journal of the Physical Society of Japan}\ }\textbf {\bibinfo
  {volume} {83}},\ \bibinfo {pages} {064003} (\bibinfo {year}
  {2014})}\BibitemShut {NoStop}%
\bibitem [{\citenamefont {Gaunt}\ \emph {et~al.}(2013)\citenamefont {Gaunt},
  \citenamefont {Schmidutz}, \citenamefont {Gotlibovych}, \citenamefont
  {Smith},\ and\ \citenamefont {Hadzibabic}}]{Gaunt2013}%
  \BibitemOpen
  \bibfield  {author} {\bibinfo {author} {\bibfnamefont {A.~L.}\ \bibnamefont
  {Gaunt}}, \bibinfo {author} {\bibfnamefont {T.~F.}\ \bibnamefont
  {Schmidutz}}, \bibinfo {author} {\bibfnamefont {I.}~\bibnamefont
  {Gotlibovych}}, \bibinfo {author} {\bibfnamefont {R.~P.}\ \bibnamefont
  {Smith}}, \ and\ \bibinfo {author} {\bibfnamefont {Z.}~\bibnamefont
  {Hadzibabic}},\ }\href {\doibase 10.1103/PhysRevLett.110.200406} {\bibfield
  {journal} {\bibinfo  {journal} {Phys. Rev. Lett.}\ }\textbf {\bibinfo
  {volume} {110}},\ \bibinfo {pages} {200406} (\bibinfo {year}
  {2013})}\BibitemShut {NoStop}%
\bibitem [{\citenamefont {Nogrette}\ \emph {et~al.}(2014)\citenamefont
  {Nogrette}, \citenamefont {Labuhn}, \citenamefont {Ravets}, \citenamefont
  {Barredo}, \citenamefont {B\'eguin}, \citenamefont {Vernier}, \citenamefont
  {Lahaye},\ and\ \citenamefont {Browaeys}}]{Nogrette2014}%
  \BibitemOpen
  \bibfield  {author} {\bibinfo {author} {\bibfnamefont {F.}~\bibnamefont
  {Nogrette}}, \bibinfo {author} {\bibfnamefont {H.}~\bibnamefont {Labuhn}},
  \bibinfo {author} {\bibfnamefont {S.}~\bibnamefont {Ravets}}, \bibinfo
  {author} {\bibfnamefont {D.}~\bibnamefont {Barredo}}, \bibinfo {author}
  {\bibfnamefont {L.}~\bibnamefont {B\'eguin}}, \bibinfo {author}
  {\bibfnamefont {A.}~\bibnamefont {Vernier}}, \bibinfo {author} {\bibfnamefont
  {T.}~\bibnamefont {Lahaye}}, \ and\ \bibinfo {author} {\bibfnamefont
  {A.}~\bibnamefont {Browaeys}},\ }\href {\doibase 10.1103/PhysRevX.4.021034}
  {\bibfield  {journal} {\bibinfo  {journal} {Phys. Rev. X}\ }\textbf {\bibinfo
  {volume} {4}},\ \bibinfo {pages} {021034} (\bibinfo {year}
  {2014})}\BibitemShut {NoStop}%
\bibitem [{\citenamefont {L\"uhmann}(2016)}]{Luhmann2016}%
  \BibitemOpen
  \bibfield  {author} {\bibinfo {author} {\bibfnamefont {D.-S.}\ \bibnamefont
  {L\"uhmann}},\ }\href {\doibase 10.1103/PhysRevA.94.011603} {\bibfield
  {journal} {\bibinfo  {journal} {Phys. Rev. A}\ }\textbf {\bibinfo {volume}
  {94}},\ \bibinfo {pages} {011603} (\bibinfo {year} {2016})}\BibitemShut
  {NoStop}%
\bibitem [{\citenamefont {Menotti}\ \emph {et~al.}(2008)\citenamefont
  {Menotti}, \citenamefont {Lewenstein}, \citenamefont {Lahaye},\ and\
  \citenamefont {Pfau}}]{Menotti2008}%
  \BibitemOpen
  \bibfield  {author} {\bibinfo {author} {\bibfnamefont {C.}~\bibnamefont
  {Menotti}}, \bibinfo {author} {\bibfnamefont {M.}~\bibnamefont {Lewenstein}},
  \bibinfo {author} {\bibfnamefont {T.}~\bibnamefont {Lahaye}}, \ and\ \bibinfo
  {author} {\bibfnamefont {T.}~\bibnamefont {Pfau}},\ }\href {\doibase
  10.1063/1.2839130} {\bibfield  {journal} {\bibinfo  {journal} {AIP Conference
  Proceedings}\ }\textbf {\bibinfo {volume} {970}},\ \bibinfo {pages} {332}
  (\bibinfo {year} {2008})}\BibitemShut {NoStop}%
\end{thebibliography}

%

\end{document}